\documentstyle[11pt,amssymb,newlfont,psfig,labelfig]{amsart}
\newcommand{\R}{\Bbb R}
\newcommand{\N}{\Bbb N}
\newcommand{\diver}{\operatorname{div}}
\newcommand{\rot}{\operatorname{curl}}
\newcommand{\grad}{\operatorname{grad}}
\newcommand{\be}[1]{\begin{equation}\label{#1}}
\renewcommand{\phi}{\varphi}
\newcommand{\eps}{\varepsilon}

\newcommand{\bml}[1]{\begin{multline}\label{#1}}
\newcommand{\bes}{\begin{equation*}}
\newcommand{\bs}{\begin{split}}
\newcommand{\bdm}{\begin{displaymath}}
\newcommand{\edm}{\end{displaymath}}

\newcommand{\di}{\partial}
\renewcommand{\phi}{\varphi}
\newtheorem{th}{Theorem}[section]
\newtheorem{lem}{Lemma}[section]

\begin{document}
\title[]{Problems on electrorheological fluid flows.}

\author[]{R.H.W. Hoppe, W.G. Litvinov\newline Lehrstuhl f\"ur Angewandte Analysis mit Schwerpunkt
Numerik \newline
Universit\"at Augsburg,
Universit\"atsstrasse, 14\newline
86159 Augsburg, 
Germany }
\date{}
\subjclass{35Q35}
\address{\newline 
E-mails: \newline
Hoppe math.uni-augsburg.de \newline
Litvinov  math.uni-augsburg.de} 

\begin{abstract}
We develop a model of an electrorheological fluid such that the fluid is 
considered as an anisotropic one with the viscosity depending on the second
 invariant of the rate of strain tensor, on the module of the vector of
 electric
field strength, and on the angle between the vectors of velocity and electric 
field. We study general problems on the flow of such fluids at nonhomogeneous
mixed boundary conditions, wherein values of velocities and surface forces 
are given on different parts of the boundary. We consider the cases where the 
viscosity function is continuous and singular, equal to infinity, when
 the second  invariant of the rate of strain tensor is equal to zero. In the
second case the problem is reduced to a variational inequality.
By using the methods of a fixed point, monotonicity, and compactness, we prove 
existence results for the problems under consideration. Some efficient methods
for numerical solution of the problems are examined.
\end{abstract}
\maketitle

\makeatletter\@addtoreset{equation}{section}\makeatother
\def\theequation{\arabic{section}.\arabic{equation}}

\section{Introduction}
\begin{figure}[htp]
\centerline{
\SetLabels
\L\B ( .18*-.08)  $(a)$\\
\L\B ( .74*-.08)  $(b)$\\
\endSetLabels
\AffixLabels{\psfig{file=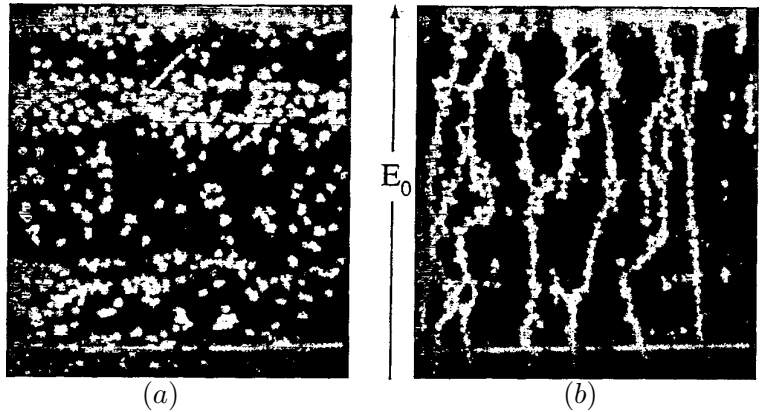,height=5cm,width=10cm}}
}
\caption{Fibrous structure formed by the electric field for alumina particles
}
\label{figFibstruc}
\end{figure}

Electrorheological fluids are smart materials which
are concentrated suspensions of polarizable particles 
in a nonconducting dielectric liquid. In moderately large electric fields, the
particles form chains along the field lines, and these chains then aggregate to 
form columns (see Fig. \ref{figFibstruc} , taken from \cite{23}). These chainlike and 
columnar structures cause dramatic changes in
the rheological properties of the suspensions. The fluids become anisotropic,
the apparent viscosity (the resistance to flow) in the direction orthogonal to
the direction of electric field abruptly increases, while the apparent viscosity
in the direction of the electric field changes not so drastically.

The chainlike and columnar structures are destroyed under the action of large 
stresses, and then the apparent viscosity of the fluid decreases and the fluid
becomes less anisotropic.

Constitutive relations for electrorheological fluids in which the stress tensor
$\sigma$ is an isotropic function of the vector of electric field strength $E$
and the rate of strain tensor $\eps$ were derived in \cite{3}, and for an 
incompressible fluid there was obtained the following equation:
\begin{equation}\label{1.1}
\sigma=-pI_1+\alpha_2 E\otimes E+\alpha_3\eps+\alpha_4\eps^2+\alpha_5(\eps E
\otimes E+E\otimes \eps E)+\alpha_6(\eps^2 E\otimes E+E\otimes \eps^2 E).
\end{equation}
Here $p$ is the pressure, $I_1$ the unit tensor, $\alpha_i$ are scalar functions
of six invariants of the tensors $\eps$, $E\otimes E$, and mixed tensors; 
$\alpha _i$ are to be determined by experiments.

 In the condition of simple 
shear flow, when the vectors of velocity $v$ and electric field $E$ are 
orthogonal and $E$ is in the plane of flow, the terms with coefficients 
$\alpha_2$, $\alpha_4$, $\alpha_5$, $\alpha_6$ give rise to two normal stresses differences (see \cite {3}). 
But these terms lead to incorrectness of the boundary value problems for
the constitutive equation \eqref{1.1}, and very restrictive conditions should be
imposed on the coefficients $\alpha_2$, $\alpha_4$, $\alpha_5$, $\alpha_6$ in
order to get an operator satisfying the conditions of coerciveness and 
monotonicity (the condition of coerciveness is almost similar to the 
Clausius-Duhem inequality following from the second law of thermodynamics, and 
the condition of monotonicity denotes that stresses increase as the rate of strains increase). 

The constitutive equation  \eqref{1.1} does not describe anisotropy of the fluid; 
in the case of simple shear flow \eqref{1.1} gives the same values of the shear 
stresses in the cases, when the vectors of velocity and electric field are 
orthogonal and parallel ($\sigma$ is an isotropic function of $E$ and $\eps$
in \eqref{1.1}).

Stationary and nonstationary mathematical problems for the special case of
\eqref{1.1} are studied in \cite{20}. It is supposed in \cite{20} that 
velocities are equal to zero everywhere on the boundary and the stress tensor 
is given by
\begin{gather}
\sigma=-pI_1+\gamma_1((1+\vert\eps\vert^2)^{\frac{k-1}2}\,-1)E\otimes E \notag\\
+(\gamma_2+\gamma_3\vert E\vert^2)(1+\vert\eps\vert^2)^{\frac{k-2}2}\,\eps+
\gamma_4(1+\vert \eps\vert^2)^{\frac{k-2}2}\,(\eps E\otimes E+E\otimes\eps E),
\label{1.2}
\end{gather}
where $\vert\eps\vert^2=\sum_{i,j=1}^n \eps_{ij}^2$, $n$ being  the dimension of
a domain of flow, $\gamma_1-\gamma_4$,  are constants, and $k$ is a function of 
$\vert E\vert^2$.

The constants $\gamma_1-\gamma_4$ and the function $k$ are determined by the approximation 
of flow curves which are obtained experimentally for different values of the 
vector of electric field $E$ (see Subsection 2.2). 
But the conditions of coerciveness and monotonicity of the
operator $-\diver(\sigma+pI_1)$ impose severe constraints on the constants 
$\gamma_1-\gamma_4$ and on the function $k$, see \cite{20}, such that with 
these restrictions one cannot obtain a good approximation of a flow curve, to 
say nothing of approximation of a set of flow curves corresponding to 
different values of $E$.

Below in Section 2, we develop a constitutive equation of electrorheological 
fluids such that a fluid is considered as a viscous one with the viscosity 
depending on the second invariant of the rate of strain tensor, on the module 
of the vector of electric field strength, and on the angle between the vectors 
of velocity and electric field  strength. This constitutive equation describes
the main peculiarities of electrorheological fluids, and it can be 
identified so that a set of flow curves corresponding to different values of 
$E$ is approximated with a high degree of accuracy, and it leads to correct
mathematical problems.

In Section 3, we present auxiliary results, and in Sections 4--8 we study problems
on stationary flow of such fluids at nonhomogeneous mixed boundary conditions.
Here we prescribe values of velocities and surface forces on different parts of 
the boundary and ignore the inertial forces. The cases where the viscosity 
function is continuous and singular, equal to infinity, when the second invariant 
of the rate of strain tensor is equal to zero, are studied. In the second case
the problem is reduced to a variational inequality.

By using the methods of a fixed point, monotonicity, and compactness we prove
existence results for the regular and singular viscosity functions. In the second
case existence results are obtained at more restrictive assumptions. Here the
singular viscosity is approximated by a continuous bounded one with a parameter 
of regularization, and a solution of the variational inequality is obtained
as a limit of the solutions of regularized problems.

Section 9 is concerned with numerical solution of the problems on stationary 
flows of electrorheological fluids with regular viscosity function. We consider
here methods of the augmented Lagrangian, Birger-Kachanov, contraction and
gradient.

In Sections 10 and 11 we study problems on flow of electrorheological fluids in
which inertial forces are taken into account. Here we consider nonhomogeneous
boundary conditions in the case that velocities are given on the whole of the 
boundary and in the case that velocities and surface forces are prescribed on 
different parts of the boundary. With some suppositions existence results are
proved.                         

\section{Constitutive equation.}
\subsection{The form of the constitutive equation.}
 
It has been found experimentally that the shear stress and accordingly the 
viscosity of electrorheological fluids depend on the shear rate, the module 
of the vector of electric field strength, and the angle between the vectors
of fluid velocity and electric fields strength (see \cite{23,1}). Thus, on the basis of experimental results we introduce the following
constitutive equation
\begin{equation}\label{2.1}
\sigma_{ij}(p,u,E)=-p\delta_{ij}+2\phi(I(u),\vert E\vert,\mu(u,E))\eps_{ij}
(u), \quad i,j=1,\dots,n, \quad n=2 \mbox{ or } 3.
\end{equation}
Here, $\sigma_{ij}(p,u,E)$ are the components of the stress tensor which 
depend on the pressure $p$, the velocity vector $u=(u_1,\dots,u_n)$ and the 
electric field strength $E=(E_1,\dots,E_n)$, $\delta_{ij}$ is the Kronecker 
delta, and 
$\eps_{ij}(u)$ are the components of the rate of strain tensor
\begin{equation}\label{2.2}
\eps_{ij}(u)=\frac12\big(\frac{\di u_i}{\di x_j}\,+\,\frac{\di u_j}{\di x_i}
\big).
\end{equation}
Moreover, $I(u)$ is the second invariant of the rate of strain tensor
\begin{equation}\label{2.3}
I(u)=\sum_{i,j=1}^n (\eps_{ij}(u))^2,
\end{equation}
and $\phi$ the viscosity function depending on $I(u)$, $\vert E\vert$ and $\mu(
u,E)$, where
\begin{equation}\label{2.4}
(\mu(u,E))(x)=\Big(\frac{u(x)}{\vert u(x)\vert},\,\,\frac{E(x)}{\vert E(x)
\vert}\Big)_{\R^n}^2\,=\,\frac{(\sum_{i=1}^n u_i(x)E_i(x))^2}{(\sum_{i=1}^n
(u_i(x))^2)(\sum_{i=1}^n(E_i(x))^2)}.
\end{equation}
So $\mu(u,E)$ is the square of the scalar product of the unit vectors $\frac
{u}{\vert u\vert}$ and $\frac E{\vert E\vert}$. The function $\mu$ is defined
by \eqref{2.4} in the case of an immovable frame of reference. If the frame
of reference moves uniformly with a constant velocity $\check u=(\check u_1,
\dots,\check u_n)$, then we set:
\begin{equation}\label{2.5}
\mu(u,E)(x)=\Big(\frac{u(x)+\check u}{\vert u(x)+\check u\vert},\,\,\frac
{E(x)}{\vert E(x)\vert}\Big)_{\R^n}^2.
\end{equation}
As the scalar product of two vectors is independent of the frame of reference,
the constitutive equation \eqref{2.1} is invariant with respect to the group
of Galilei transformations of the frame of reference that are represented as
a product of time-independent translations, rotations and uniform motions.

It is obvious that $\mu(u,E)(x)\in[0,1]$, and  for fixed
$y_1,y_2\in\R_+$, where $\R_+=\{z\in\R,\,\,\,z\ge 0\}$, the function $y_3\to
\phi(y_1,y_2,y_3)$ reaches its maximum at $y_3=0$ and its minimum at $y_3=1$ 
when the
vectors $u(x)+\check u$ and $E$ are  correspondingly orthogonal and parallel.

 The function $\mu$ defined by \eqref{2.4}, \eqref{2.5} is not specified at
$E=0$ and at $u=0$, and there does not exist an extension by continuity 
to the values of $u=0$ and $E=0$. However, at $E=0$ there is no influence of 
the electric field. Therefore,
\begin{equation}\label{2.6}
\phi(y_1,0,y_3)=\tilde\phi(y_1),\qquad  y_3\in[0,1],
\end{equation}
and the function $\mu(u,E)$ need not be specified at $E=0$. Likewise, in case 
that the 
measure of the set of points $x$ at which $u(x)=0$ is zero, the function $\mu$ 
need not also be specified at $u=0$. But in the general case we should specify 
$\mu$ for all values of $u$. Because of this we assume that the  function $\mu$
is defined as follows:
\begin{equation}\label{2.7}
\mu(u,E)(x)=\Big(\frac{\alpha\tilde I+u(x)+\check u}{\alpha\sqrt n+\vert u(x)+
\check u\vert},\,\,\frac{E(x)}{\vert E(x)\vert}\Big)_{\R^n}^2,
\end{equation}
where $\tilde I$ denotes  a vector with components equal to one, and $\alpha$ 
is a small
positive constant. If $u(x)\ne 0$ almost everywhere in $\Omega$, we may choose
$\alpha=0$.

\subsection{Assumptions on the viscosity function.}

\begin{figure}[htp]
\centerline{
\SetLabels
\L\B ( -.03*-.05)  $0$\\
\L\B ( 1.01*-.015)  $\gamma$\\
\L\B ( -.0*1.01)  $\tau$\\
\L\B ( -.05*.4)  $\tau_0$\\
\L\B ( .75*-.05)  $\gamma_1$\\
\L\B ( .04*-.05)  $\gamma_0$\\
\L\B ( .72*.33)  $1$\\
\L\B ( .72*.53)  $2$\\
\L\B ( .72*.79)  $3$\\
\L\B ( .72*.98)  $4$\\
\endSetLabels
\AffixLabels{\psfig{file=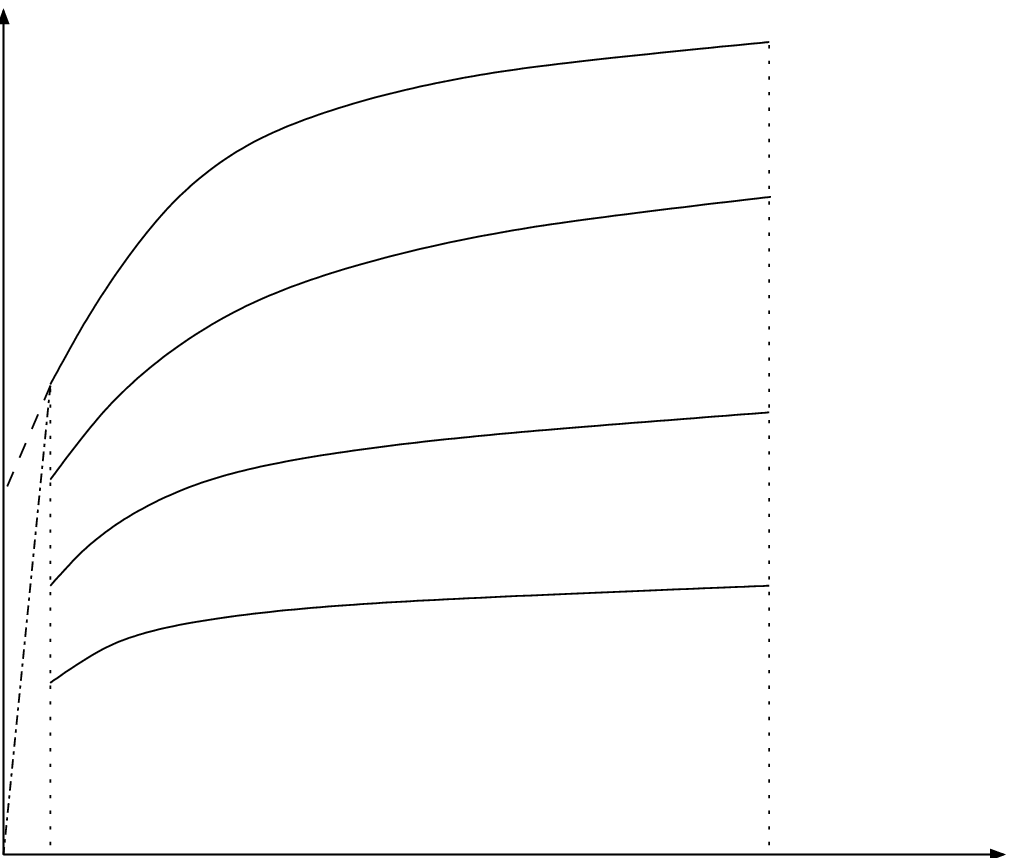,height=6cm,width=8.5cm}}
}
\caption{ }
\label{figFlow}
\end{figure}

Flow curves of electrorheological fluids obtained experimentally for 
$\mu(u,E)=0$ have the form as displayed in Fig. \ref{figFlow} (cf.,e.g.,\cite{1}).
 
These curves define the relationship between the shear stress $\tau=\sigma_{12}$
and the
shear rate $\gamma=\eps_{12}(u)=\frac12\,\frac{d u_1}{d x_2}$ for a flow that is
close to simple shear flow. Line 1 is the flow curve for $\vert E\vert=0$, and
lines 2--4 represent the flow curve for increasing  $\vert E\vert$. 

Flow curves are obtained in some region, say $\gamma_0\le\gamma\le\gamma_1$,
 $\gamma_0>0$. Experimental results for small $\gamma$ are not precise, and
one has to extend the flow curves to $\R_+$. It is customary to extend flow
curves by straight lines over the region $\gamma_1<\gamma<\infty$. One can 
prolong flow curves in $[0,\gamma_0)$ such that either $\tau=\tau_0$ for 
$\gamma=0$ or $\tau=0$ for $\gamma=0$ (see the dash and dot-dash lines in 
Fig.2).

The viscosity $\eta(\gamma,E)$ of the fluid is determined as 
\begin{equation}\label{2.9}
\eta(\gamma,E)=\frac12\,\frac{\tau}{\gamma},
\end{equation}
and it is defined by the approximation of the lines 1--4 extended to $\R_+$. 
Generalizing \eqref{2.9} to an arbitrary flow, we take
\begin{equation}\label{2.10}
\gamma=\Big(\frac12 I(u)\Big)^{\frac12},\quad \phi(I(u),\vert E\vert,\,\,
\mu(u,E))=\eta\Big(\Big(\frac12 I(u)\Big)^{\frac12},E\Big).
\end{equation}
If the flow curve is extended by the straight line $\tau=c_1+c_2\gamma$, 
$\gamma\in(\gamma_1,\infty)\,$, we obtain
\begin{gather}
\phi(I(u),\vert E\vert,\mu(u,E))=
\begin{cases}
\phi_1(I(u),\vert E\vert,\mu(u,E)),\quad &I(u)\in [0,2\gamma_1^2],\\
\frac12(c_2+c_1(\frac 12 I(u))^{-\frac12}), \quad &I(u)\in (2\gamma_1^2,\infty).
\end{cases}
\label{2.11}
\end{gather}
Here, the coefficients $c_1$ and $c_2$ depend on $\vert E\vert$ and $\mu(u,E)$.
The viscosity function $\phi$ is continuous in $\R_+^2\times[0,1]$, if the 
flow curve is extended in $[0,\gamma_0]$ by the dot-dash line, and it has the 
form
\begin{equation}\label{2.12}
\phi(I(u),\vert E\vert,\mu(u,E))=\frac{b(\vert E\vert,\mu(u,E))}{I(u)^
{\frac12}}+\psi(I(u),\vert E\vert,\mu(u,E))
\end{equation}
with $b(\vert E\vert,\mu(u,E))=2^{-\frac12}\,\tau_0$, if the flow curve is extended
by the dash line in $[0,\gamma_0]$, $\psi$ being a function continuous in
$\R_+^2\times [0,1]$.

We note that if $\psi(I(u),\vert E\vert,\mu(u,E))=b_1(\vert E\vert,\mu(u,E))$,
then
\eqref{2.12} is the viscosity function of an extended Bingham electrorheological 
fluid. 

In the case that the flow curve is extended in $[0,\gamma_0]$ by the dot-dash
line, the viscosity function can be written as follows:
\begin{equation}\label{2.12a}
\phi(I(u),\vert E\vert,\mu(u,E))=b(\vert E\vert,\mu(u,E))(\lambda+I(u))^
{-\frac12}+\psi(I(u),\vert E\vert,\mu(u,E)),
\end{equation}
where $\lambda$ is a small positive parameter. Obviously, for $\lambda=0$
the function $\phi$ defined by \eqref{2.12a} is the same as the one defined by 
\eqref{2.12}. Moreover, for  $I(u)=0$  we have
\begin{align}
&\phi(0,\vert E\vert,\mu(u,E))=\infty \mbox{ for \eqref{2.12}}, \notag\\
&\phi(0,\vert E\vert,\mu(u,E))=b(\vert E\vert,\mu(u,E))\lambda^{-\frac12}
+\psi(0,\vert E\vert,\mu(u,E))\mbox{ for \eqref{2.12a}}.\notag
\end{align}
Flow problems for fluids with a constitutive equation \eqref{2.12}
reduce to the solution of variational inequalities. Such problems are 
considerably more complicated than problems for fluids with finite viscosity,
in particular, for fluids with  a
constitutive equation as given by \eqref{2.12a}. From a physical point of
view, \eqref{2.12a} with a finite, but possibly large viscosity for $I(u)=0$ 
seems to be more reasonable than \eqref{2.12}.

We will study problems in the case that $\phi$ is a continuous bounded function
of its arguments and in the case that $\phi$ is singular and the singular part
of the function $\phi$ is equal to $b(\vert E\vert,\mu(u,E))I(u)^{-\frac12}$. 
In the first case we assume that $\phi$ satisfies one of the following 
conditions (C1), (C2), (C3):
\begin{description}
\item[(C1)]
               $\phi:(y_1,y_2,y_3)\to\phi(y_1,y_2,y_3)$ is a function 
               continuous in $\R_+^2\times[0,1]$, and for an arbitrarily
               fixed $(y_2,y_3)\in\R_+\times[0,1]$ the function $\phi
               (.,y_2,y_3):y_1\to\phi(y_1,y_2,y_3)$ is continuously
               differentiable in $\R_+$, and the following inequalities 
               hold:
\end{description}
\begin{gather}
a_2\ge\phi(y_1,y_2,y_3)\ge a_1 \label{2.16}\\
\phi(y_1,y_2,y_3)+2\,\frac{\di\phi}{\di y_1}\,(y_1,y_2,y_3)y_1\ge a_3
\label{2.17}\\
\Big\vert \frac{\di\phi}{\di y_1}\,(y_1,y_2,y_3)\Big\vert y_1\le a_4,
\label{2.18}
\end{gather}
where $a_i,1\le i\le 4$, are positive numbers.
\begin{description} 
\item[(C2)]
           $\phi:(y_1,y_2,y_3)\to\phi(y_1,y_2,y_3)$ is a function continuous
           in $\R_+^2\times[0,1]$, and for an arbitrarily fixed $(y_2,y_3)
           \in\R_+\times[0,1]$, \eqref{2.16} and the following inequality hold:
\end{description}
\begin{gather}
[\phi(z_1^2,y_2,y_3)z_1-\phi(z_2^2,y_2,y_3)z_2](z_1-z_2)\ge a_3(z_1-z_2)^2
 \quad \forall (z_1,z_2)\in\R_+^2.\label{2.19}
\end{gather}

\begin{description}
\item[(C3)]
           $\phi:(y_1,y_2,y_3)\to\phi(y_1,y_2,y_3)$ is a function continuous
           in $\R_+^2\times[0,1]$, and for an arbitrarily fixed $(y_2,y_3)\in
           \R_+\times[0,1]$, \eqref{2.16} holds and the function $z\to\phi(
           z^2,y_2,y_3)z$ is strictly increasing in $\R_+$, i.e., the 
           conditions
           $z_1,z_2\in\R_+$, $z_1>z_2$ imply $\phi(z_1^2,y_2,y_3)z_1>\phi
          (z_2^2,y_2,y_3)z_2$.
\end{description}

Let us dwell on the physical sense of these inequalities. \eqref{2.16} 
indicates that the viscosity is bounded from below and from above by positive
constants.
The inequality \eqref{2.17} implies that for fixed values of $\vert E\vert$
and $\mu(u,E)$ the derivative of the function $I(v)\to G(v)$ is positive, where
$G(v)$ is the second invariant of the stress deviator
\begin{equation}
G(v)=4[\phi(I(v),\vert E\vert,\,\,\,\mu(u,E))]^2I(v).  \notag
\end{equation}
This means that in case of simple shear flow the shear stress
increases with  increasing shear rate. \eqref{2.18} is a restriction
on $\frac{\di\phi}{\di y_1}$ for large values of $y_1$. 
These inequalities are natural from a physical point of view.

The assumptions (C2) and (C3) indicate that in case of simple shear flow,
the shear stress must increase with increasing  shear rate. 

Moreover, let $\lambda(z)=\phi(z^2,y_2,y_3)z$. Assume that the function 
$\lambda$ is continuously differentiable in $\R_+$. It follows from (C2) that
$\frac{d\lambda}{d z}\ge a_3$ for all $z\in\R_+$. Calculating $\frac{d\lambda}
{d z}(z)$, we obtain \eqref{2.17} from \eqref{2.19}. On the contrary, 
\eqref{2.19} follows from \eqref{2.17}. The assumption (C3) indicates that 
$\frac{d\lambda}{d z}(z)>0$ for all $z\in\R_+$ or, equivalently 
\begin{equation}
\phi(y_1,y_2,y_3)+2\frac{\di\phi}{\di y_1}(y_1,y_2,y_3)y_1>0, \quad  
 y_1\in\R_+. \notag
\end{equation}
The following assumption is concerned with the function (coefficient) 
$b$ in \eqref{2.12}, \eqref{2.12a}:
\begin{description} 
\item[(C4)]
           $b:y_1,y_2\to b(y_1,y_2)$ is a function continuous
           in $\R_+\times[0,1]$ and in addition
\begin{equation}\label{2.20}
 0\le b(y_1,y_2)\le a_5, \quad   (y_1,y_2)\in\R_+\times[0,1],
\end{equation}
$a_5$ being a positive number.
\end{description}

Generally, the continuous function $\phi$ is expressible in the form
\begin{equation}\label{2.21a}
\phi(I(u),\vert E\vert,\mu(u,E))=\sum_{i=1}^{m} e_i(\vert E\vert,\mu
(u,E))\beta_i(I(u)).
\end{equation}
Polynomials or splines can be used to represent the functions $\beta_i$. 
The flow curves  obtained for various values of $E$ can be approximated with an
arbitrary accuracy. We note that \eqref{2.21a} may also be used
for an identification of the function $\psi$.

In the general case,  the coefficients $e_i$ as well as the viscosity function 
$\phi$ depend on the temperature  and these coefficients  can be determined 
by an approximation of the corresponding flow curves.

\subsection{General problems.}

Fig. \ref{figElectric} below gives an example of an electrorheological fluid flow.
\begin{figure}[htp]
\centerline{
\SetLabels
\L\B ( .1*-.05)  $S_1=S\setminus\overline{S_2}$\\
\L\B ( .1*.6)  $S_2$\\
\L\B ( .13*.45)  $\Omega_1$\\
\L\B ( .62*.45)  $\Omega_3$\\
\L\B ( .3*.65)  $\Gamma_0$\\
\L\B ( .3*.22)  $\Gamma_1$\\
\L\B ( .4*.7)  $\Omega_2$\\
\L\B ( .95*.44)  $\Delta U(t)$\\
\L\B ( -.08*.45)  $F$\\
\L\B ( .88*.45)  $\hat u$\\
\endSetLabels
\AffixLabels{\psfig{file=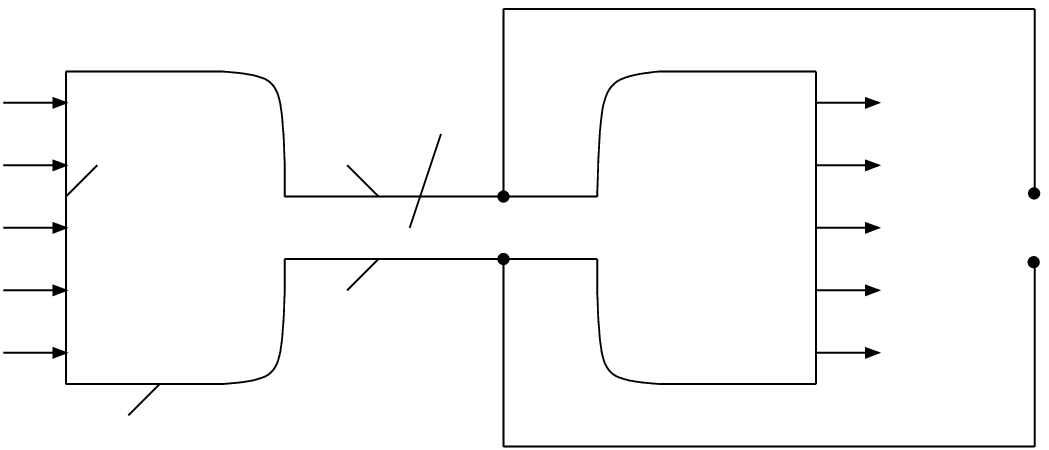,height=3.5cm,width=8.5cm}}
}
\caption{}
\label{figElectric}
\end{figure}

Here, the domain $\Omega$ of fluid flow consists of three parts $\Omega_1$,
$\Omega_2$, $\Omega_3$. A fluid flows from the part $\Omega_1$ across $\Omega_
2$ in the part $\Omega_3$. Electrodes are placed on parts $\Gamma_0$ and $\Gamma
_1$ of the boundary of $\Omega_2$, and an electric field $E$ is generated by
applying voltages  $\bigtriangleup U(t)$ to electrodes at time $t$.
Generally it may be a $k$ pairs of electrodes and voltages $\bigtriangleup U_i
(t)$ are applied to $i$-th pair of electrodes,  $i=1,\dots,k$. 
The boundary $S$ of the domain $\Omega$ consists of two 
parts $S_1$ and $S_2$. Surface forces $F=(F_1,\dots,F_n)$ act on $S_2$, and the
distribution of velocities $\hat u=(\hat u_1,\dots,\hat u_n)$ is given
on $S_1$.

The equations of motion and the incompressibility condition read as follows:
\begin{gather}
\rho\Big(\frac{\di u_i}{\di t}+u_j\,\frac{\di u_i}{\di x_j}\Big)+\frac{\di p}
{\di x_i}-2\,\frac{\di}{\di x_j}\Big[\phi(I(u),\vert E\vert,\mu(u,E))\eps_
{ij}(u)\Big]  \notag\\
=K_i\quad\mbox{in}\quad Q=\Omega\times(0,T),\quad i=1,\dots,n, \label{2.21}\\
\diver\,u=\sum_{i=1}^n \frac{\di u_i}{\di x_i}=0 \quad\mbox{in}\quad Q.
\label{2.22}
\end{gather}
Here, $K_i$ are the components of the volume force vector $K$, $\rho$ is the 
density, $T$ a positive constant. In \eqref{2.21} and below Einstein's 
convention on summation over repeated index is applied.

We assume that $\Omega$ is a bounded domain in $\R^n$, $n=2$ or $3$.
Suppose that $S_1$ and $S_2$ are open subsets of $S$ such that $S=\overline
S_1\cup\overline S_2$ and $S_1\cap S_2=\emptyset$. The boundary and initial
conditions are the following:
\begin{gather}
u\big\vert_{S_1\times(0,T)}=\hat u,  \label{2.23}\\
[-p\delta_{ij}+2\phi(I(u),\vert E\vert,\mu(u,E))\eps_{ij}(u)]\nu_j\big\vert_
{S_2\times(0,T)}=F_i \quad  i=1,\dots,n, \label{2.24}\\
u(.,0)=u^0 \quad \mbox{in}\quad \Omega. \label{2.25}
\end{gather}
Here, $F_i$ and $\nu_j$ are the components of the vector of surface force 
$F=(F_1,\dots,F_n)$ and  the unit outward normal 
$\nu=(\nu_1,\dots,\nu_n)$ to $S$, respectively.

We consider Maxwells equations in the following form (see e.g. \cite{21}):
\begin{gather}
\rot E+\frac 1c\frac{\di B}{\di t}\,=0, \qquad \diver B=0, \notag\\
\rot H-\frac 1c\frac{\di D}{\di t}\,=0, \qquad \diver D=0. \label{2.26}
\end{gather}
Here $E$ is the electric field, $B$ the magnetic induction, $D$ the electric
displacement,  $H$ the magnetic field, $c$ the speed of light. One can 
assume that
\begin{equation}\label{2.27}
D=\epsilon E, \qquad B=\mu H,
\end{equation}
where $\epsilon$ is the dielectric permittivity, $\mu$ the magnetic permeability.

Since electrorheological fluids are dielectrics the magnetic field $H$ can be 
neglected. Then \eqref{2.26}, \eqref{2.27} give the following relations
\begin{gather}
\rot E=0, \label{2.28}\\
\diver (\epsilon E)=0. \label{2.29}
\end{gather}
It follows from \eqref{2.28} that there exists a function of potential $\theta$
such that
\begin{equation}\label{2.30}
E=-\grad\theta, 
\end{equation}
and \eqref{2.29} implies
\begin{equation}\label{2.31}
\diver(\epsilon\grad\theta)=0.
\end{equation}
The boundary conditions are the following:
\begin{gather}
\theta=U_i(t) \quad\mbox{on}\quad \Gamma_i, \quad i=1,\dots,k, \label{2.32}\\
\theta=0 \quad \mbox{on}\quad \Gamma_{i0}, \label{2.33}\\
\nu\cdot\epsilon\,\grad\theta=0 \quad\mbox{on}\quad S\backslash(\bigcup_
{i=1}^k\overline{(\Gamma_i\cup\Gamma_{i0}))}. \label{2.34}
\end{gather}
Here $\Gamma_i$ and $\Gamma_{i0}$ are the surfaces of the i-th control and
null electrodes respectively, and it is supposed that $\Gamma_i$, $\Gamma_{i0}$
are open subset of $S$. We assume
\begin{equation}\label{2.35}
\epsilon\in L_\infty(\Omega), \quad e_1\le\epsilon\le e_2\quad \mbox{a.e. in }\Omega, 
\end{equation}
$e_1$, $e_2$ being positive constants. Suppose also that
\begin{equation}\label{2.36}
U_i(t)\in H_{00}^{\frac 12}(\Gamma_i), \quad t\in [0,T], \quad i=1,\dots,k,
\end{equation}
where 
\begin{equation}\label{2.37}
H_{00}^{\frac 12}(\Gamma_i), \quad\{\psi\vert\psi=v\big\vert_{\Gamma_i}, \quad
v\in H^1(\Omega),\quad v\big\vert_{S\backslash\overline{\Gamma_i}}=0\}.
\end{equation}
The space $H_{00}^{\frac 12}(\Gamma_i)$ is normed by
\begin{equation}\label{2.38}
\|\psi\|_{H_{00}^{\frac 12}(\Gamma_i)}=\inf\{\|v\|_{H^1(\Omega)}, \quad 
v\in H^1(\Omega), \quad v\big\vert_{\Gamma_i}=\psi, \quad 
v\big\vert_{S\backslash\overline{\Gamma_i}}=0\}. 
\end{equation}
Let $\check\theta$ be a function such that
\begin{equation}\label{2.39}
\check \theta \in H^1(\Omega), \quad \check\theta\big\vert_{\Gamma_i}\,=U_i(t),
\quad \check\theta\big\vert_{\Gamma_{i0}}\,=0, \quad i=1,\dots,k.
\end{equation}
Define a space $\tilde V$ and a bilinear form $a:H^1(\Omega)\times\tilde V\to\R$
as follows:
\begin{gather}
\tilde V=\{ v\in H^1(\Omega), \qquad v\Big\vert_{\cup_{i=1}^k(\Gamma_i\cup
\Gamma_{i0})}=0  \}, \label{2.40}\\
a(v,h)=\int_\Omega\,\epsilon\frac{\di v_i}{\di x_i}\,\frac{\di h}{\di x_i}\,dx,
\qquad v\in H^1(\Omega),\qquad h\in\tilde V. \label{2.41}
\end{gather}
Consider the problem: find $u$ satisfying
\begin{gather}
u\in \tilde V \notag\\
a(u,h)=-a(\check\theta,h) \qquad h\in\tilde V. \label{2.42}
\end{gather}
By use of Green's formula it can be seen that, if $u$ is a solution of problem
\eqref{2.42}, then the function $\theta=u+\check\theta$ is a solution of \eqref
{2.31}--\eqref{2.34} in the sense of distributions. On the contrary, if $\theta$
is a solution of problem \eqref{2.31}--\eqref{2.34}, then $u=\theta-\check\theta$
is a solution of the problem \eqref{2.42}. Therefore, the function $\theta=u+
\check\theta$ is a generalized solution of problem \eqref{2.31}--\eqref{2.34}.

\begin{th}
Suppose the conditions \eqref{2.35}, \eqref{2.36} are satisfied. Then there 
exists a unique generalized solution $\theta$ of problem \eqref{2.31}--\eqref
{2.34}, and the function $\theta$ is represented in  the form $\theta=\check
\theta+u$, where $\check\theta$ satisfies \eqref{2.39} and $u$ is the solution
of problem \eqref{2.42}.
\end{th}
{\bf Proof}. By virtue of \eqref{2.35} the bilinear form a is continuous and
coercive in $\tilde V$. Therefore there exists a unique solution $u$ of problem
\eqref{2.42},
and the function $\theta=\check\theta+u$ is a generalized solution of problem
\eqref{2.31}--\eqref{2.34}.

Let $\theta_1=\check\theta_1+u_1$ and $\theta_2=\check\theta_2+u_2$ be two generalized
solutions of problem \eqref{2.31}--\eqref{2.34}, where $\check\theta_1$ and 
$\check\theta_2$ satisfy \eqref{2.39}, and $u_1,\,u_2\in\tilde V$. Then $\theta
_1-\theta_2=w\in\tilde V$, and \eqref{2.42} implies $a(w,w)=0$. Therefore
$w=0$.
$\blacksquare$

The functions of volume force $K$ and surface force $F$ in \eqref{2.21} and
\eqref{2.24} are represented in the form
\begin{equation}\label{2.43}
K=\tilde K+K_e, \qquad F=\tilde F+F_e,
\end{equation}
where $\tilde K$ and $\tilde F$ are the main volume and surface forces, $K_e$ and
$F_e$ are volume and surface forces generated by the vector of electric field 
$E$. Considering electrorheological fluid as a liquid dielectric we present the 
stress tensor $\sigma_e=\{\sigma_{eik}\}_{i,k=1}^n$ induced by electric field
as follows (see \cite{21}):
\begin{equation}\label{2.44}
\sigma_{eik}=-\frac{\vert E\vert^2}{8\pi}\big(\epsilon-\rho
\frac{\di\epsilon}{\di\rho}\big)\delta_{ik}+\frac{\epsilon}{4\pi}E_i E_k.
\end{equation}
Taking into account \eqref{2.28}, we obtain the following formula for the 
vector of volume force
\begin{gather}
K_e=(K_{e1},\dots,K_{en}), \notag\\
K_{ei}=\frac{\di\sigma_{eij}}{\di x_j}=-\frac{\vert E\vert^2}{8\pi}\,\frac
{\di\epsilon}{\di x_i}\,+\,\frac 1{8\pi}\,\frac{\di}{\di x_i}\Big(\vert E
\vert^2\rho\frac{\di\epsilon}{\di\rho}\Big), \quad i=1,\dots,n. \label{2.45}
\end{gather}
The vector of surface forces is given by
\begin{equation}\label{2.46}
F_e=(F_{e1},\dots,F_{en}), \qquad F_{ei}=\sigma_{eik}\nu_k.
\end{equation}
Thus the systems \eqref{2.21}--\eqref{2.25} and \eqref{2.31}--\eqref{2.34}
are separated, so one can first solve quasi-static system \eqref{2.31}--\eqref
{2.34} and then solve the problem \eqref{2.21}--\eqref{2.25}, \eqref{2.43}.

\section{Auxiliary results.}
Let $\Omega$ be a bounded domain in $\R^n$ with a Lipschitz continuous
boundary $S$, $n=2$ or 3. Let $S_1$ be an open non-empty subset of $S$.
We consider the following spaces:
\begin{gather}
X=\{u\vert u\in H^1(\Omega)^n,\,\,u\vert_{S_1}=0\}, \label{3.1}\\
V=\{u\vert u\in X,\,\,\diver u=0\}. \label{3.2}
\end{gather}
By means of Korn's inequality, the expression
\begin{equation}\label{3.3}
\|u\|_X=\left(\int_\Omega I(u)dx\right)^{\frac12}
\end{equation} 
defines a norm on $X$ and $V$ being equivalent to the norm of $H^1
(\Omega)^n$.

Everywhere below we use the following notations: If $Y$ is a normed space,
we denote by $Y^*$ the dual of $Y$, and by $(f,h)$ the duality between $Y^*$
and $Y$, where $f\in Y^*$, $h\in Y$. In particular, if $f\in L_2(\Omega)$ or
$f\in L_2(\Omega)^n$, then $(f,h)$ is the scalar product in $L_2(\Omega)$ or
in $L_2(\Omega)^n$, respectively. The sign $\rightharpoonup$ denotes weak
convergence in a Banach space.

We further consider three functions $\tilde v,v_1,v_2$ such that
\begin{gather}
\tilde v\in H^1(\Omega)^n,\quad v_1\in L_2(\Omega),\quad v_1(x)\ge 0\mbox{ a.
e. in }\Omega,\notag\\
v_2\in L_\infty(\Omega),\quad v_2(x)\in[0,1]\mbox{ a.e. in }\Omega.\label{3.4}
\end{gather}
We set $v=(\tilde v,v_1,v_2)$ and define an operator $L_v:X\to X^*$ as follows:
\begin{equation}\label{3.5}
(L_v(u),h)=2\int_\Omega\phi(I(u+\tilde v),v_1,v_2)\eps_{ij}(u+\tilde v)
\eps_{ij}(h)dx \quad u,h\in X.
\end{equation}
\begin{lem}
Suppose the conditions $(C1)$ and \eqref{3.4} are satisfied. Then the 
following inequalities hold
\begin{gather}
(L_v(u)-L_v(w),u-w)\ge\mu_1\|u-w\|_X^2,    \qquad  u,w\in X, \label{3.6}\\
\|L_v(u)-L_v(w)\|_{X^*}\le\mu_2\|u-w\|_X,  \qquad  u,w\in X, \label{3.7}
\end{gather}
where
\begin{equation}\label{3.8}
\mu_1=\min(2a_1,2a_3),\quad \mu_2=2a_2+\,4a_4.
\end{equation}
\end{lem}
{\it Proof}. Let $u,w$ be arbitrarily fixed functions in $X$ and
\begin{equation}\label{3.9}
h=u-w.
\end{equation}
We introduce the function $\gamma$ as follows:
\begin{gather}
\gamma(t)=\int_\Omega \phi(I(\tilde v+w+th),v_1,v_2)\eps_{ij}(\tilde v+w+th)
\eps_{ij}(e)dx, \quad t\in[0,1],\quad e\in X. \label{3.10}
\end{gather}
It is obvious that
\begin{equation}\label{3.11}
\gamma(1)-\gamma(0)=\frac12(L_v(u)-L_v(w),e).
\end{equation}
By classical analysis it follows that $\gamma$ is differentiable at any point 
$t\in(0,1)$. Therefore
\begin{equation}\label{3.12}
\gamma(1)=\gamma(0)+\frac{d\gamma}{dt}(\xi), \quad \xi\in(0,1),
\end{equation}
where
\begin{gather}
\frac{d\gamma}{dt}(\xi)=\int_\Omega[\phi(I(\tilde v+w+\xi h),v_1,v_2)\eps_{ij}
(h)\eps_{ij}(e)  \notag\\
+2\frac{\di\phi}{\di y_1}\,(I(\tilde v+w+\xi h),v_1,v_2)\eps_{km}(\tilde v+
w+\xi h)\eps_{km}(h)\eps_{ij}(\tilde v+w+\xi h)\eps_{ij}(e)]dx.
 \label{3.13}
\end{gather}
Taking note of the inequality 
\begin{equation}\notag
\vert\eps_{ij}(\tilde v+w+\xi h)\eps_{ij}(h)\vert\le I(\tilde v+w+\xi h)^
{\frac12} I(h)^{\frac12}
\end{equation}
and \eqref{2.16}, \eqref{2.18} we get \eqref{3.7} as a direct consequence of
\eqref{3.9}--\eqref{3.13}.

Define the function $g$ as follows:
\begin{equation}
g(\alpha,x)=\begin{cases}
\frac{\di\phi}{\di y_1}(\alpha,v_1(x),v_2(x)), &\mbox{ if }\frac{\di\phi}
{\di y_1}(\alpha,v_1(x),v_2(x))<0,\\
0, &\mbox{ if }\frac{\di\phi}{\di y_1}(\alpha,v_1(x),v_2(x))\ge 0,
\end{cases}\notag
\end{equation}
where $\alpha\in\R_+$, $x\in\Omega$.

Then, taking $e=h$ in \eqref{3.13} and applying \eqref{2.17} we get
\begin{gather}
\frac{d\gamma}{dt}(\xi)=\int_\Omega[\phi(I(\tilde v+w+\xi h),v_1,v_2)I(h)
\notag\\
+2\frac{\di\phi}{\di y_1}(I(\tilde v+w+\xi h),v_1,v_2)(\eps_{ij}(\tilde v
+w+\xi h)\eps_{ij}(h))^2]dx\ge \min(a_1,a_3)\|h\|_X^2 \label{3.14}
\end{gather}
and \eqref{3.6} follows from \eqref{3.14}.
$\blacksquare$
\begin{lem}
Let the function $\phi$ satisfy  condition $(C2)$ and assume that \eqref{3.4}
holds true. Then the operator $L_v$ is a continuous mapping from $X$ into $X^*$
and
\begin{equation}\label{3.15}
(L_v(u)-L_v(w),u-w)\ge 2a_3\int_\Omega [I(u+\tilde v)^{\frac12}-I(w+\tilde v)
^{\frac12}]^2 dx, \quad  u,w\in X.
\end{equation}
Moreover,
\begin{equation}\label{3.16}
(L_v(u)-L_v(w),u-w)=0 \Longleftrightarrow u=w.
\end{equation} 
\end{lem}

{\it Proof}. We set $u^1=u+\tilde v$, $w^1=w+\tilde v$. Taking into account that 
$[\eps_{ij}(u^1)\eps_{ij}(w^1)]\le I(u^1)^{\frac12}I(w^1)^{\frac12}$, we
obtain
\begin{gather}
(L_v(u)-L_v(w),u-w)=(L_v(u)-L_v(w),u^1-w^1)  \notag\\
=2\int_\Omega[\phi(I(u^1),v_1,v_2)I(u_1)+\phi(I(w^1),v_1,v_2)I(w^1) \notag\\
-\phi(I(u^1),v_1,v_2)\eps_{ij}(u^1)\eps_{ij}(w^1)-\phi(I(w^1),v_1,v_2)\eps_
{ij}(w^1)\eps_{ij}(u^1)]dx  \notag\\
\ge 2\int_\Omega[\phi(I(u^1),v_1,v_2)I(u^1)^{\frac12}-\phi(I(w^1),v_1,v_2)
I(w^1)^{\frac12}][I(u^1)^{\frac12}-I(w^1)^{\frac12}]dx  \label{3.17}
\end{gather}
Observing \eqref{2.19}, \eqref{3.17} yields \eqref{3.15}. Now, assume
\begin{equation}\label{3.18}
(L_v(u)-L_v(w),u-w)=0.
\end{equation}
Then, by \eqref{3.15} we have
\begin{equation}\label{3.19}
I(u+\tilde v)=I(w+\tilde v)\mbox{ a.e. in }\Omega,\quad \phi(I(u^1),v_1,v_2)
=\phi(I(w^1),v_1,v_2)\mbox{ a.e. in }\Omega.
\end{equation}
Taking \eqref{2.16}, \eqref{3.5}, \eqref{3.18},\eqref{3.19} into account 
we get $\|u-w\|_X=0$. 

The continuity of the mapping $L_v$ follows from the 
continuity of the Nemytskii operator (see \cite{6}, and also \cite{7}, Lemma
8.2 Chapter 2).
$\blacksquare$
\begin{lem}
Assume that $(C3)$ is satisfied and \eqref{3.4} holds true. Then 
\begin{equation}\label{3.20}
(L_v(u)-L_v(w),u-w)\ge 0, \quad u,\,w\in X.
\end{equation}
Moreover, \eqref{3.16} is valid, and the operator $L_v$ is a continuous mapping 
from $X$ into $X^*$.
\end{lem}
{\it Proof.} Indeed, \eqref{3.20} follows from (C3) and \eqref{3.17}. Assume 
that \eqref{3.18} is valid. Then (C3) and \eqref{3.17} imply \eqref{3.19}, and 
by \eqref{2.16} we obtain $u=w$.
$\blacksquare$

Define the set $U$ as follows
\begin{equation}\label{3.21}
U=\{h\in L_\infty(\Omega), \quad 0\le h(x)\le a_5 \mbox{ a.e. in }\Omega\},
\end{equation}
and let $\tilde v\in H^1(\Omega)^n$. For a given constant $\lambda>0$ define
an operator $L_\lambda:U\times X\to X^*$ as follows:
\begin{equation}\label{3.22}
(L_\lambda(h,u),w)=\int_\Omega h(\lambda+I(\tilde v+u))^{-\frac12}
\,\eps_{ij}(\tilde v+u)\eps_{ij}(w)dx.
\end{equation}
\begin{lem}
For an arbitrary $\lambda>0$ and an arbitrarily fixed $h\in U$ the 
following inequalities hold:
\begin{gather}
(L_\lambda(h,u_1)-L_\lambda(h,u_2),u_1-u_2)\ge 0, \quad  u_1,u_2\in X,
\label{3.23}\\
\|L_\lambda(h,u_1)-L_\lambda(h,u_2)\|_{X^*}\le\alpha\|u_1-u_2\|_X,\quad 
\alpha=2a_5\lambda^{-\frac12}, \label{3.24}\\
\|L_\lambda(h,u)\|_{X^*}\le \alpha_1,\quad \alpha_1=\left(\int_\Omega h^2\,dx
\right)^{\frac12}. \label{3.25}
\end{gather}
Moreover, the conditions
\begin{gather}
\{h_m\}\subset U,\quad h_m\to h \mbox{ a.e. in }\Omega, \notag \\
u_m\to u \mbox{ in } X,\quad  u_m\to u\mbox{ a.e. in }\Omega,\quad\frac
{\di u_{mi}}{\di x_j}\to\frac{\di u_i}{\di x_j} \mbox{ a.e. in }\Omega, 
\quad i,j=1,\dots,n,
\label{3.26}
\end{gather}
imply
\begin{equation}\label{3.27}
L_\lambda(h_m,u_m)\to L_\lambda(h,u) \mbox{ in } X^*.
\end{equation}
\end{lem}
{\it Proof}. The viscosity function associated with  the operator 
$L_\lambda(h,.):u\to L_\lambda(h,u)$ has the form
\begin{equation}\label{3.28}
\phi(y)=\frac12 h(\lambda+y)^{-\frac12},\quad\quad y\in\R_+, 
\end{equation}
where $y$ plays the role of the second invariant of the rate of strain tensor.
We have
\begin{gather}
\phi(y)+2\frac{d\phi}{dy}(y)y 
=\frac12 h(\lambda+y)^{-\frac12}[1-(\lambda+y)^{-1}y]>0, \quad
 y\in \R_+, \quad \lambda>0. \label{3.29} 
\end{gather}
The left-hand side of \eqref{3.29} represents the derivative of the function
$g:z\to g(z)=\phi(z^2)z$, $z^2=y$. Therefore, the function $g$ is increasing,
and \eqref{3.23} follows  from the proof of Lemma 3.3. 

By \eqref{3.21}, \eqref{3.22} we obtain
\begin{equation}\notag
\vert(L_\lambda(h,u),w)\vert\le\int_\Omega h(\lambda+I(\tilde v+u))^{-\frac
12}\,I(\tilde v+u)^{\frac12}\,I(w)^{\frac12}\,dx\le\left(\int_\Omega h^2\,dx
\right)^{\frac12}\|w\|_X,
\end{equation}
which readily gives \eqref{3.25}. Moreover, \eqref{3.21} and \eqref{3.28} yield
\begin{equation}\label{3.30}
 \phi(y) \le\frac12 a_5\lambda^{-\frac12},\,\,\,
\Big\vert\frac{d\phi}{dy}(y)\Big\vert y\le \frac 14 a_5\lambda^{-\frac12} ,
\qquad  y\in\R_+,
\end{equation}
and hence, observing Lemma 3.1 we get \eqref{3.24}.

Assuming \eqref{3.26}, it follows that
\begin{gather}
\|L_\lambda(h_m,u_m)-L_\lambda(h,u)\|_{X^*}\le\|L_\lambda(h_m,u_m)-L_\lambda
(h_m,u)\|_{X^*} \notag\\
+\|L_\lambda(h_m,u)-L_\lambda(h,u)|_{X^*}. \label{3.31}
\end{gather}
Further,
\begin{gather}
(L_\lambda(h_m,u_m)-L_\lambda(h_m,u),w)=\int_\Omega h_m[(\lambda+I(\tilde v+
u_m))^{-\frac12}\eps_{ij}(u_m-u)\eps_{ij}(w)\notag\\
+((\lambda+I(\tilde v+u_m))^{-\frac12}-(\lambda+I(\tilde v+u))^{-\frac12})
\eps_{ij}(\tilde v+u)\eps_{ij}(w)]dx,\notag
\end{gather}
whence
\begin{gather}
\|L_\lambda(h_m,u_m)-L_\lambda(h_m,u)\|_{X^*} 
\le\left[\int_\Omega h_m^2(\lambda+I(\tilde v+u_m))^{-1}I(u_m-u)dx
\right]^{\frac12}\notag\\
+\left\{\int_\Omega h_m^2[(\lambda+I(\tilde v+u_m))^{-\frac12}-
(\lambda+I(\tilde v+u))^{-\frac12}]^2 I(\tilde v+u)dx\right\}^{\frac12}.
\label{3.32}
\end{gather}
Obviously, the first term of the right-hand side in \eqref{3.32} tends to 
zero. By \eqref{3.26} we have $I(\tilde v+u_m)\to I(\tilde v+u)$
a.e. in $\Omega$, and by the Lebesgue theorem we obtain that the second term
of the right-hand side in \eqref{3.32} tends to zero. The second term of the 
right-hand side in \eqref{3.31} also tends to zero. Thus \eqref{3.27} is 
satisfied, and the lemma is proven.
$\blacksquare$

\begin{lem}
Let $\Omega$ be a bounded domain in $\R^n$, $n=2$ or $3$ with a Lipschitz
continuous boundary $S$, and let the operator $B\in\cal L(X,L_2(\Omega))$ be
defined as follows:
\begin{equation}\label{3.33}
Bu=\diver u.
\end{equation}
Then, the $\inf$-$\sup$ condition
\begin{equation}\label{3.34}
\inf_{\mu\in L_2(\Omega)}\,\,\sup_{v\in X}\,\,\frac{(Bv,\mu)}{\|v\|_X\,\|\mu
\|_{L_2(\Omega)}}\,\,\ge \beta_1>0
\end{equation}
holds true. The operator $B$ is an isomorphism from $V^\bot$  onto $L_2(\Omega)$, where
$V^\bot$ is the orthogonal complement of $V$ in $X$, and the operator $B^*$ that is
adjoint to $B$, is an isomorphism from $L_2(\Omega)$ onto the polar  set 
\begin{equation}\label{3.35}
V^0=\{f\in X^*,\,(f,u)=0, \quad u\in V\}.
\end{equation}
Moreover,
\begin{gather}
\|B^{-1}\|_{\cal L(L_2(\Omega),V^\bot)}\le \frac1{\beta_1}, \label{3.36}\\
\|(B^*)^{-1}\|_{\cal L(V^0,L_2(\Omega))}\le \frac1{\beta_1}. \label{3.37}
\end{gather}
\end{lem}
For a proof see in \cite{8}. Lemma 3.5 is a generalization of the inf-sup
condition in case that the operator div acts in the subspace 
$H^1_0(\Omega)$ (see \cite{9}). This result was first established in an 
equivalent form by Ladyzhenskaya and Solonnikov in \cite{10}.

Let $\{X_m\}_{m=1}^\infty$, $\{N_m\}_{m=1}^\infty$ be sequences of 
finite-dimensional subspaces in $X$ and $L_2(\Omega)$, respectively, such that
\begin{gather}
 \lim_{m\to\infty}\,\,\inf_{z\in X_m}\,\|u-z\|_X=0, \quad u\in X, \label{3.38}\\
\lim_{m\to\infty}\,\,\inf_{y\in N_m}\,\|w-y\|_{L_2(\Omega)}=0, \quad 
 w\in L_2(\Omega). \label{3.39}
\end{gather}
Define the operators $B_m\in\cal L(X_m,N_m^*)$ as follows:
\begin{equation}\label{3.40}
(B_mu,\mu)=\int_\Omega \mu\,\diver u\,dx, \qquad u\in X_m, \quad \mu\in N_m,
\end{equation}
and let $B_m^*\in\cal L(N_m,X_m^*)$ be the adjoint operator of $B_m$ with
$(B_mu,\mu)=(u,B_m^*\mu)$ for all $u\in X_m$ and all $\mu\in N_m$.

We introduce the spaces $V_m$ and $V_m^0$ by
\begin{gather}
V_m=\{u\in X_m,\,\,\,(B_mu,\mu)=0,\quad \mu\in N_m\}, \label{3.41}\\
V_m^0=\{q\in X_m^*,\,\,\,(q,u)=0, \quad u\in V_m\}.   \label{3.42} 
\end{gather}
The following Lemma is valid (see \cite{8}).
\begin{lem}
Let $\{X_m\}_{m=1}^\infty$, $\{N_m\}_{m=1}^\infty$ be sequences of 
finite-dimensional subspaces in $X$ and $L_2(\Omega)$ and assume that the 
discrete $\inf-\sup$ condition (LBB condition)
\begin{equation}\label{3.43}
\inf_{\mu\in N_m}\,\sup_{u\in X_m}\,\frac{(B_mu,\mu)}{\|u\|_X\|\mu\|_{L_2
(\Omega)}}\,\ge\beta>0,\quad m\in\N
\end{equation}
holds true.
Then the operator $B_m^*$ is an isomorphism from $N_m$ onto $V_m^0$, and the
operator $B_m$ is an isomorphism from $V_m^\bot$ onto $N_m^*$, where $V_m^
\bot$ is an orthogonal complement of $V_m$ in $X_m$. Moreover, 
\begin{equation}\label{3.44}
\|(B_m^*)^{-1}\|_{\cal L(V_m^0,N_m})\le\frac1{\beta}, \quad \|B_m^{-1}\|_
{\cal L(N_m^*,V_m^\bot)}\le \frac1{\beta},\quad  m\in\N.
\end{equation}
\end{lem}
Consider a functional $\Psi:U\times X\to\R_+$ of the form
\begin{equation}\label{3.45}
\Psi(h,u)=\int_\Omega hI(u)^{\frac 12}dx \qquad h\in U,\quad u\in X,
\end{equation}
where $U$ is as in \eqref{3.21}.

\begin{lem}
For an arbitrarily fixed $h\in U$ the functional $\Psi(h,.):u\to\Psi(h,u)$ is
continuous in $X$ and the conditions      
\begin{equation}\label{3.46}
\{h_m\}\subset U, \quad h_m\to h \text{ a.e. in } \Omega,\quad u_m\rightharpoonup 
u \text{ in }X
\end{equation}
imply
\begin{equation}\label{3.47}
\lim\inf\Psi(h_m,u_m)\ge\Psi(h,u).
\end{equation}
\end{lem}
Here and below the sign $\rightharpoonup$ designates the weak convergence.

{\it Proof of the Lemma 3.7}. Let $u_m\to u$ in $X$. We have
\begin{equation}\notag
\int_\Omega hI(u_m-u)^{\frac12}\,dx\le\Big(\int_\Omega h^2\,dx\Big)^{\frac12}
\Big(\int_\Omega I(u_m-u)dx\Big)^{\frac12}.
\end{equation}
Therefore,
\begin{equation}\label{3.50a}
\lim\int_\Omega hI(u_m-u)^{\frac12}\,dx=0,
\end{equation}
and
\begin{equation}\notag
\int_\Omega hI(u_m-u)^{\frac12}\,dx\ge\Big\vert\int_\Omega hI(u_m)^{\frac12} 
\,dx-\int_\Omega hI(u)^{\frac12}\,dx\Big\vert.
\end{equation}
Consequently, $\lim\Psi(h,u_m)=\Psi(h,u)$.

Let $\alpha\in [0,1]$, $u,v\in X$. Then
\begin{gather}
I(\alpha u+(1-\alpha)v)=I(\alpha u)+2\alpha(1-\alpha)\sum_{i,j=1}^n \eps_{ij}
(u)\eps_{ij}(v) \notag\\
+I((1-\alpha)v)\le[\alpha I(u)^{\frac12}+(1-\alpha)I(v)^{\frac12}]^2. 
\label{3.48} 
\end{gather}
Therefore,
\begin{gather}
\Psi(h,\alpha u+(1-\alpha)v)=\int_\Omega h I(\alpha u+(1-\alpha)v)^{\frac 12}
dx\le \alpha\Psi(h,u)+(1-\alpha)\Psi(h,v), \label{3.49}
\end{gather}
which shows that
\begin{equation}\label{3.50}
\Psi(h,.):u\to\Psi(h,u) \mbox{ is a convex functional in }X.
\end{equation}
Let now \eqref{3.46} be fulfilled. We have
\begin{gather}
\Psi(h_m,u_m)=\int_\Omega[hI(u_m)^{\frac12}+(h_m-h)I(u_m)^{\frac 12}]\,dx,
\label{3.51}\\
\Big\vert\int_\Omega(h_m-h)I(u_m)^{\frac12}\,dx\Big\vert\le\|h_m-h\|_{L_2
(\Omega)}\|u_m\|_X. \label{3.52}
\end{gather}
In view of \eqref{3.46}, the right-hand side of \eqref{3.52} tends to zero as $m\to
\infty$. Hence \eqref{3.50}, \eqref{3.51}, and the continuity of the 
functional $\Psi(h,.)$ imply
\begin{equation}\label{3.53}
\lim\inf\Psi(h_m,u_m)=\lim\inf\Psi(h,u_m)\ge\Psi(h,u),
\end{equation}
and the lemma is proved.
$\blacksquare$

{\bf Remark 3.1}. Assume that
\begin{equation}\label{3.54}
h_1\le h(x)\le h_2\quad \mbox{ a.e. in } \Omega,
\end{equation}
where $h_1$, $h_2$ are positive constants. Then the expression
\begin{equation}\label{3.55}
\int_\Omega h I(u)^{\frac 12}\,dx=\|u\|_h
\end{equation}
defines a norm on  $X$. Note that this norm 
is not equivalent to the norm of the space $W_1^1(\Omega)^n$. However, 
\begin{equation}\label{3.56}
\|u\|_p=\left(\int_\Omega hI(u)^{\frac p2}\,dx\right)^{\frac 1p}
\end{equation}
is a norm on $X$, which is  equivalent to the norm 
of $W_p^1(\Omega)^n$ for $p>1$ (cf., e.g., \cite{12}).

\section{The stationary problem.}

We consider stationary flow problems of electrorheological fluids 
under the Stokes approximation, i.e., we ignore inertial forces. Such an
approach
is reasonable, because the viscosities of electrorheological fluids are large,
and the inertial terms have a small impact. We 
deal with the following problem: find a pair of functions $u$, $p$ satisfying
\begin{gather}
\frac{\di p}{\di x_i}-2\,\frac{\di}{\di x_j}\left[\phi(I(u),\vert E\vert,
\mu(u,E))\eps_{ij}(u)\right]=K_i \quad \text{ in }\Omega,\quad i=1,\dots,n, 
\label{4.1}\\
\diver\,u=0\quad \text{in }\Omega, \label{4.2}\\
u\Big\vert_{S_1}=\hat u,  \label{4.3}\\
[-p\delta_{ij}+2\phi(I(u),\vert E\vert,\mu(u,E))\eps_{ij}(u)]\nu_j\Big\vert
_{S_2}=F_i, \quad i=1,\dots,n.\label{4.4}
\end{gather}
We assume that
\begin{equation}\label{4.5}
\hat u\in H^{\frac12}(S_1)^n.
\end{equation}
Then there exists a function $\tilde u$  such that
\begin{equation}\label{4.6}
\tilde u\in H^1(\Omega)^n,\quad \tilde u\Big\vert_{S_1}=\hat u,\quad\diver\,
\tilde u=0.
\end{equation}
Suppose also
\begin{equation}\label{4.7}
K=(K_1,\dots,K_n)\in L_2(\Omega)^n,\quad F=(F_1,\dots,F_n)\in L_2(S_2)^n.
\end{equation}
In line with \eqref{2.12}, we choose the viscosity function $\phi$ of the 
following form:
\begin{equation}\label{4.8}
\phi(I(u),\vert E\vert,\mu(u,E))=\frac{b(\vert E\vert,\mu(u,E))}{I(u)^
{\frac12}}+\psi(I(u),\vert E\vert,\mu(u,E)),
\end{equation}
where $\psi$ is a function satisfying one out of the conditions (C1), (C2), 
(C3) with $\phi$ replaced by $\psi$, and $b$ satisfies (C4).
We refer to the fluid with the viscosity function $\phi$ defined by \eqref{4.8} 
as a generalized Bingham electrorheological fluid.

Define a functional $J$ on the set $X\times X$ and an operator $L:X\to X^*$
as follows:
\begin{gather}
J(v,h)=2\int_\Omega b(\vert E\vert,\mu(\tilde u+v,E))I(\tilde u+h)^{\frac12}
dx, \quad v,h\in X.\label{4.9}\\
(L(v),h)=2\int_\Omega\psi(I(\tilde u+v),\vert E\vert,\mu(\tilde u+v,E))
\eps_{ij} (\tilde u+v)\eps_{ij}(h)dx, \quad v,h\in X. \label{4.10}
\end{gather}
We use the notations
\begin{equation}\label{4.11}
(K,h)=\int_\Omega K_i h_i dx,\quad (F,h)=\int_{S_2}F_i h_i ds, \qquad h\in X.
\end{equation}
The following assertion holds.

\begin{th}
Let $\Omega$ be a bounded domain in $\R^n$, $n=2$ or $3$, with a Lipschitz 
continuous boundary $S$. Assume \eqref{4.6}, \eqref{4.7} are satisfied and 
$(u,p)$ with $u=\tilde u+v$ is a regular solution of \eqref{4.1}
--\eqref{4.4}, where the viscosity function $\phi$ is defined by \eqref{4.8}
with $\psi$ meeting one out of the conditions $(C1)$, $(C2)$, $(C3)$  
($\phi$ replaced by $\psi$) and $b$ satisfying $(C4)$. Then 
\begin{gather}
v\in V, \label{4.12}\\
J(v,h)-J(v,v)+(L(v),h-v)\ge(K+F,h-v), \quad  h\in V. \label{4.13}
\end{gather}
\end{th}
{\it Proof}. Let $h=(h_1,\dots,h_n)\in V$. We multiply \eqref{4.1} with $h_i-v_i$,
sum over $i$ and integrate over $\Omega$. By  Green's formula and
\eqref{4.4}, \eqref{4.8}, we obtain
\begin{gather}
2\int_\Omega b(\vert E\vert,\mu(\tilde u+v,E))I(\tilde u+v)^{-\frac12}\eps
_{ij}(\tilde u+v)\eps_{ij}(h-v)dx \notag\\
+(L(v),h-v)=(K+F,h-v), \qquad h\in V. \label{4.14}
\end{gather}
We use the relations
\begin{equation}\label{4.15}
\eps_{ij}(h-v)=\eps_{ij}(\tilde u+h)-\eps_{ij}(\tilde u+v), \quad
\eps_{ij}(\tilde u+v)\eps_{ij}(\tilde u+h)\le I(\tilde u+v)^{\frac12}
I(\tilde u+h)^{\frac12},  
\end{equation}
so that the first addendum of the left-hand side of \eqref{4.14} is majorized
by $J(v,h)-J(v,v)$. Then, \eqref{4.14} implies \eqref{4.13}, and the theorem is
proved.
$\blacksquare$

Let $v$ be a solution of the problem \eqref{4.12}, \eqref{4.13} such
that
\begin{equation}\label{4.17}
I(\tilde u+v)\ne 0 \text{ a.e. in }\Omega.
\end{equation}
We replace $h$ in \eqref{4.13} by $v+\lambda h$, $\lambda>0$, from which
\begin{equation}\label{4.18}
\lambda^{-1}[J(v,v+\lambda h)-J(v,v)]+(L(v),h)\ge(K+F,h).
\end{equation}
For $\lambda\to 0$ we get
\begin{equation}\label{4.19}
\left(\frac{\di J}{\di h}(v,v),h\right)+(L(v),h)\ge (K+F,h),\quad h\in V,
\end{equation}
where $\frac{\di J}{\di h}(v,v)$ is the partial G\^ateaux derivative of the 
functional $J$ with respect to the second argument
\begin{equation}\label{4.20}
\left(\frac{\di J}{\di h}(v,v),h\right)=2\int_\Omega b(\vert E\vert,\mu
(\tilde u+v,E))I(\tilde u+v)^{-\frac12}\eps_{ij}(\tilde u+v)\eps_{ij}(h)dx.
\end{equation}
Replacing $h$ by $-h$ in \eqref{4.19} we obtain
\begin{equation}\label{4.21}
\left(\frac{\di J}{\di h}(v,v),h\right)+(L(v),h)=(K+F,h),\quad h\in V,
\end{equation}
and Lemma 3.5 gives
\begin{equation}\label{4.22}
\frac{\di J}{\di h}(v,v)+L(v)-K-F=B^*p, \quad p\in L_2(\Omega),
\end{equation}
that is
\begin{equation}\label{4.23}
\left(\frac{\di J}{\di h}(v,v),h\right)+(L(v),h)-(B^*p,h)=(K+F,h) \quad
\forall h\in X.
\end{equation}
It follows from \eqref{4.23} that the pair $(u,p)$ with $u=\tilde u+v$ is a 
solution of \eqref{4.1}--\eqref{4.4} in the sense of 
distributions. Thus, we have proved the following:

{\bf Remark 4.1} If $v$ is a solution of \eqref{4.12}, 
\eqref{4.13} that satisfies \eqref{4.17}, then there exists a function
$p\in L_2(\Omega)$ such that the pair $(u,p)$ with $u=\tilde u+v$ is a solution
of \eqref{4.1}--\eqref{4.4} in the distributional sense. In view 
of this and Theorem 4.1 it is reasonable to refer to the function 
$u=\tilde u+v$ as a generalized solution of \eqref{4.1}--\eqref{4.4}.

\section{Problem for the fluid with constitutive equation \eqref{2.12a}.}
\subsection{Existence theorem.}
We define the following functional on $X\times X$:
\begin{equation}\label{5.1}
J_\lambda(v,h)=2\int_\Omega b(\vert E\vert,\mu(\tilde u+
v,E))(\lambda+I(\tilde u+h))^{\frac12}\,dx,\quad \lambda>0.
\end{equation}
Obviously, $J_\lambda(v,h)=J(v,h)$ for $\lambda=0$. Note that the functional
$J_\lambda$ is G\^ateaux differentiable in $X$ with respect to the second 
argument for $\lambda>0$, but not for  $\lambda=0$.

The partial G\^ateaux derivative $\frac{\di J_\lambda}{\di h}$ is given by
\begin{gather}
\Big(\frac{\di J_\lambda}{\di h}(v,h),w\Big)=2\int_\Omega b(\vert E\vert,\mu
(\tilde u+v,E))(\lambda+I(\tilde u+h))^{-\frac12} 
\eps_{ij}(\tilde u+h)\eps_{ij}(w)dx, \notag\\
  v,h,w\in X,\quad \lambda>0. 
\label{5.2}
\end{gather}
Consider the following problem: find $v_\lambda$ such that
\begin{gather}
v_\lambda\in V, \label{5.3}\\
\Big(\frac{\di J_\lambda}{\di h}(v_\lambda,v_\lambda),w\Big)+(L(v_\lambda),w)
=(K+F,w), \quad  w\in V. \label{5.4}
\end{gather}
Lemma 3.5 implies that if $v_\lambda$ is a solution of \eqref{5.3},
\eqref{5.4}, then there exists a function $p_\lambda$ such that the pair 
($v_\lambda$, $p_\lambda$) is a solution of the following problem:
\begin{gather}
(v_\lambda,p_\lambda)\in X\times L_2(\Omega), \label{5.5}\\
\Big(\frac{\di J_\lambda}{\di h}(v_\lambda,v_\lambda),w\Big)+(L(v_\lambda),w)-
(B^* p_\lambda,w)=(K+F,w), \quad  w\in X, \label{5.6}\\
(Bv_\lambda,q)=0, \qquad  q\in L_2(\Omega). \label{5.7}
\end{gather}
We remark that \eqref{5.5}--\eqref{5.7} represent  the flow of an
electrorheological fluid with the constitutive equation \eqref{2.12a}.
We seek an approximate solution of the problem \eqref{5.5}--\eqref{5.7} of the
form
\begin{gather}
(v_m,p_m)\in X_m\times N_m,\label{5.8}\\
\Big(\frac{\di J_\lambda}{\di h}(v_m,v_m),w\Big)+(L(v_m),w)-(B^*_m p_m,w)=
(K+F,w), \quad  w\in X_m, \label{5.9}\\
(B_m v_m,q)=0, \qquad  q\in N_m, \label{5.10}
\end{gather}
where $X_m$ and $N_m$ are finite dimensional subspaces in $X$ and $L_2,
(\Omega)$, respectively, and $B_m$ is defined as in \eqref{3.40}.

\begin{th}
Suppose that the conditions $(C4)$, \eqref{4.6}, \eqref{4.7} are satisfied and the 
function $\psi$ meets one of the conditions $(C1)$, $(C2)$, $(C3)$ ($\phi$
replaced by $\psi$). Let $\{X_m\}$, $\{N_m\}$ be sequences of finite-dimensional
subspaces 
in $X$ and $L_2(\Omega)$, respectively, such that \eqref{3.38}, \eqref{3.39}, 
\eqref{3.43} hold and
\begin{equation}\label{5.11}
X_m\subset X_{m+1}, \quad N_m\subset N_{m+1},\quad  m\in\N. 
\end{equation}
Then, for an arbitrary $\lambda>0$ there exists a solution $(v_\lambda$, 
$p_\lambda)$ of \eqref{5.5}--\eqref{5.7}. Moreover, for $m\in\N$ and 
$\lambda>0$ there exists a solution 
$(v_m$, $p_m)$ of \eqref{5.8}--\eqref{5.10}, and a subsequence 
$\{(v_k, p_k)\}$ can be extracted from the sequence $\{(v_m,p_m)\}$ such that 
$v_k\rightharpoonup v_\lambda$ in $X$, $p_k\rightharpoonup p_\lambda$ in $L_2
(\Omega)$.
\end{th}
{\it Proof}. It follows from \eqref{3.41}, \eqref{5.8}--\eqref{5.10} that 
$v_m$ is a solution of the problem
\begin{equation}\label{5.12}
v_m\in V_m, \quad \Big(\frac{\di J_\lambda}{\di h}(v_m,v_m),w\Big)+(L(v_m),w)
=(K+F,w), \quad  w\in V_m. 
\end{equation}
Taking  \eqref{2.20} into account we obtain
\begin{gather}
\Big\vert\Big(\frac{\di J_\lambda}{\di h}(e,e),e\Big)\Big\vert=2\Big\vert\int_
\Omega b(\vert E\vert,\mu(\tilde u+e,E))\frac{\eps_{ij}(\tilde u+e)\eps_{ij}
(e)}{(\lambda+I(\tilde u+e))^{\frac12}}dx\Big\vert \notag\\
\le 2\int_\Omega b(\vert E\vert,\mu(\tilde u+e,E))I(e)^{\frac12}\,dx
\le c_1\|e\|_X, \quad \lambda>0,\quad  e\in X,\label{5.13}
\end{gather}
where
\begin{equation}\label{5.14}
c_1=2a_5(\mbox{mes }\Omega)^{\frac12}.
\end{equation}
By \eqref{2.16}, \eqref{4.6}, \eqref{4.7}, \eqref{4.10}, and \eqref{5.13}, for an
arbitrary $e\in X$ we get
\begin{equation}\label{5.16}
z(e)=\Big(\frac{\di J_\lambda}{\di h}(e,e),e\Big)+(L(e),e)-(K+F,e)
\ge 2a_1\|e\|_X^2-c\|e\|_X, \quad e\in X, \quad \lambda>0,
\end{equation}
giving  $\quad z(e)\ge 0\quad$ for $\|e\|_X\ge r=\frac c{2a_1}$.

From the corollary of Brouwer's fixed point theorem (cf.\cite{13}) it follows
that there exists a solution of \eqref{5.12} with  
\begin{equation}\label{5.17}
\|v_m\|_X\le r, \qquad \|L(v_m)\|_{X^*}\le c_2, \quad  m\in\N,
\end{equation}
where the second inequality  follows from \eqref{2.16} and \eqref{4.6}.
For an arbitrary $f\in X^*$ we denote by $Gf$ the restriction of $f$ to
$X_m$. Then $Gf\in X_m^*$, and by \eqref{3.42}, \eqref{5.12} we obtain
\begin{equation}\label{5.18}
G\Big(\frac{\di J_\lambda}{\di h}(v_m,v_m)+L(v_m)-K-F\Big)\in V_m^0.
\end{equation}
Therefore, there exists a unique $p_m\in N_m$ (see Lemma 3.6) such that
\begin{equation}\label{5.19}
B_m^*p_m=G\Big(\frac{\di J_\lambda}{\di h}(v_m,v_m)+L(v_m)-K-F\Big).
\end{equation}
Thus the pair ($v_m$, $p_m$) is a solution of  \eqref{5.8}--\eqref{5.10}. 
Due to \eqref{2.20}, \eqref{4.7}, \eqref{5.17} and Lemmas 3.4, 3.6 we get
\begin{equation}\label{5.20}
\|p_m\|_{L_2(\Omega)}\le c, \quad  m\in\N.
\end{equation}
By \eqref{5.17}, \eqref{5.20} we can extract a subsequence $\{v_\eta,p_\eta\}$
such that
\begin{gather}
v_\eta\rightharpoonup v_0 \quad \mbox{ in } X,\label{5.21}\\
v_\eta\to v_0 \mbox{ in } L_2(\Omega)^n \mbox{ and a.e. in } \Omega,
\label{5.22}\\
p_\eta\rightharpoonup p_0 \quad \mbox{ in } L_2(\Omega), \label{5.23}\\
L(v_\eta)\rightharpoonup \chi  \quad \mbox{ in } X^*, \label{5.24}\\
\frac{\di J_\lambda}{\di h}(v_\eta,v_\eta)\rightharpoonup\chi_1 \quad\mbox{in }
X^*. \label{5.25}
\end{gather}

Let $\eta_0$ be a fixed positive number and $w\in X_{\eta_0}$, $q\in N_
{\eta_0}$. Observing \eqref{5.21}, \eqref{5.23}--\eqref{5.25} we pass to the limit in
\eqref{5.9}, \eqref{5.10} with $m$ replaced by $\eta$, and obtain
\begin{gather}
(\chi_1+\chi-B^*p_0,w)=(K+F,w), \quad w\in X_{\eta_0},\notag\\
\int_\Omega q\diver v_0\,dx=0, \quad  q\in N_{\eta_0}.\label{5.26}
\end{gather}
Since $\eta_0$ is an arbitrary positive integer, by 
\eqref{3.38}, \eqref{3.39} 
\begin{gather}
\chi_1+\chi-B^*p_0=K+F, \label{5.27}\\
\diver v_0=0. \label{5.28}
\end{gather}
We present the operator $L(v)$ in the form
\begin{equation}\label{5.29}
L(v)=L(v,v),
\end{equation}
where the operator ($v,w)\to L(v,w)$ is considered as a mapping of $X\times
X$ into $X^*$ according to
\begin{equation}\label{5.30}
(L(v,w),h)=2\int_\Omega \psi(I(\tilde u+w),\vert E\vert,\mu(\tilde u+v,E))
\eps_{ij}(\tilde u+w)\eps_{ij}(h)dx.
\end{equation} 
We get
\begin{equation}\label{5.31}
X_\eta(w)=\left(\frac{\di J_\lambda}{\di h}(v_\eta,v_\eta)+L(v_\eta,v_\eta)-
\frac{\di J_\lambda}{\di h}(v_\eta,w)-L(v_\eta,w),v_\eta-w\right),\quad
w\in X.
\end{equation}
Lemmas 3.3, 3.4 imply
\begin{equation}\label{5.32}
X_\eta(w)\ge 0, \quad \eta\in\N, \quad  w\in X.
\end{equation}
We have
\begin{gather}
\Big\|\frac{\di J_\lambda}{\di h}(v_\eta,w)-\frac{\di J_\lambda}{\di h}(v_0,w)
\Big\|_{X^*} \notag\\
\le 2\Big[\int_\Omega[b(\vert E\vert,\mu(\tilde u+v_\eta,E))-b(\vert E\vert,\mu
(\tilde u+v_0,E))]^2\,dx\Big]^{\frac12}.\label{5.33}
\end{gather}
\eqref{2.20}, \eqref{5.22}, \eqref{5.33} and the Lebesgue theorem give
\begin{equation}\label{5.34}
\frac{\di J_\lambda}{\di h}(v_\eta,w)\to \frac{\di J_\lambda}{\di h}(v_0,w)
\quad \text{in } X^*.
\end{equation}
Likewise we obtain
\begin{equation}\label{5.35}
L(v_\eta,w)\to L(v_0,w) \quad\text{in } X^*.
\end{equation}
Taking into account that $(B_\eta v_\eta,p_\eta)=0$, by \eqref{5.9}, 
\eqref{5.21}, \eqref{5.23} we obtain
\begin{equation}\label{5.36}
\Big(\frac{\di J_\lambda}{\di h}(v_\eta,v_\eta)+L(v_\eta),v_\eta\Big)=(K+F,
v_\eta)\to(K+F,v_0), 
\end{equation}
and
\begin{equation}\label{5.37}
\lim\left(\frac{\di J_\lambda}{\di h}(v_\eta,v_\eta)+L(v_\eta),w\right)-
(B^*p_0,w)=(K+F,w), \quad w\in X.
\end{equation}
Observing \eqref{5.34}--\eqref{5.37} and passing to the limit in \eqref{5.31}, 
by \eqref{5.28}, \eqref{5.32} we get
\begin{equation}\label{5.38}
\left(K+F-\frac{\di J_\lambda}{\di h}(v_0,w)-L(v_0,w)+B^*p_0,v_0-w\right)\ge 0,
\quad  w\in X.
\end{equation}
We choose $w=v_0-\gamma h$, $\gamma>0$, $h\in X$, and consider $\gamma\to 0$.
Then, Lemmas 3.3, 3.4 give
\begin{equation}\label{5.39}
\left(K+F-\frac{\di J_\lambda}{\di h}(v_0,v_0)-L(v_0,v_0)+B^* p_0,h\right)\ge 0.
\end{equation}
This inequality holds for any $h\in X$. Therefore, replacing $h$ by $-h$ shows
that equality holds true. Consequently, the pair ($v_\lambda$, $p_\lambda$) 
with $v_\lambda=v_0$ and $p_\lambda=p_0$ solves \eqref{5.5}--\eqref{5.7}. The 
theorem is proved.
$\blacksquare$

\subsection{On the uniqueness of the solution.}
Let $v_\lambda$, $w_\lambda$ be two solutions of \eqref{5.5}--\eqref{5.7} and
\begin{equation}\label{5.39a}
e=w_\lambda-\,v_\lambda.
\end{equation}
Define a function $\eta$ as follows:
\begin{equation}\label{5.40}
\eta(t)=\Big(\frac{\di J_\lambda}{\di h}(v_\lambda+t\,e,\,v_\lambda+t\,e),e
\Big)+(L(v_\lambda+t\,e),e),\quad t\in [0,1].
\end{equation}
It follows from \eqref{5.6} that
\begin{equation}\label{5.41}
\eta(1)-\eta(0)=0.
\end{equation}
Assume that the function $\mu$ is defined by \eqref{2.7}. Suppose also that the
functions $b$ and $\psi$ are continuously differentiable and in addition
\begin{equation}\label{5.42}
\Big\vert \frac{\di\psi}{\di y_3}(y_1,\vert E(x)\vert,y_3)\Big\vert y_1^
{\frac12}\le \tilde c,\quad (y_1,y_3)\in \R_+\times[0,1],\quad x\in\Omega. 
\end{equation}
Note that \eqref{5.42} is a restriction on the behaviour of the function
$\frac{\di\psi}{\di y_3}$ at large values of $y_1$. Under the above conditions 
the function $\eta$ is differentiable, and we have
\begin{equation}\label{5.43}
\eta(1)-\eta(0)=\frac{d\eta}{dt}(\xi), \quad \xi\in(0,1).
\end{equation}
Here
\begin{equation}\notag
\frac{d\eta}{dt}(\xi)=\sum_{i=1}^4\,\gamma_i(\xi),
\end{equation}
where
\begin{gather}
\gamma_1(\xi)=2\int_\Omega\,\frac{\di b}{\di y_2}(\vert E\vert,\mu(\tilde u+
v_\lambda+\xi e,E))\,f_\xi\,(\lambda+I(\tilde u+v_\lambda+\xi e))^{-\frac12}
\notag\\
\times\eps_{ij}\,(\tilde u+v_\lambda+\xi e)\,\eps_{ij}\,(e)\,dx. 
\label{5.44}\\
f_\xi=2\Big(\frac{\alpha\tilde I+\tilde u+v_\lambda+\xi e}{\alpha\sqrt{n}+
\vert\tilde u+v_\lambda+\xi e\vert},\,\frac E{\vert E\vert}\Big)_{\R^n}\,\,
\Big[\Big(\frac e{\alpha\sqrt{n}+\vert\tilde u+v_\lambda+\xi e\vert},\,\frac E
{\vert E\vert}\Big)_{\R^n} \notag\\
-(\alpha\sqrt{n}+\vert\tilde u+v_\lambda+\xi e\vert)
^{-2}\,\,\vert\tilde u+v_\lambda+\xi e\vert^{-1}  \notag\\
\times((\tilde u_i+v_{\lambda i}+\xi e_i)e_i)\Big(\alpha\tilde I+\tilde u+v_
\lambda+\xi e,\,\,\frac E{\vert E\vert}\Big)_{\R^n}\Big], 
\label{5.45}\\
\gamma_2(\xi)=2\int_\Omega b(\vert E\vert,\mu(\tilde u+v_\lambda+\xi e,E))
[-(\lambda+I(\tilde u+v_\lambda+\xi e))^{-\frac32}\notag\\
\times\eps_{km}\,(\tilde u+v_\lambda+\xi e)\,\eps_{km}(e)\,
\eps_{ij}(\tilde u+v_\lambda+\xi e)\,\eps_{ij}(e)+
(\lambda+I(\tilde u+v_\lambda+\xi e))^{-\frac12}\,\,I(e)]\,dx,
\label{5.46}\\
\gamma_3(\xi)=2\int_\Omega \frac{\di\psi}{\di y_3}(I(\tilde u+v_\lambda+
\xi e),\vert E\vert,\mu(\tilde u+v_\lambda+\xi e,E))f_\xi \notag\\
\times\eps_{ij}(\tilde u+v_\lambda+\xi e)\eps_{ij}(e)\,dx, 
\label{5.47}\\
\gamma_4(\xi)=2\int_\Omega[\psi(I(\tilde u+v_\lambda+\xi e),\vert E\vert,\mu
(\tilde u+v_\lambda+\xi e,E))I(e) \notag\\
+2\frac{\di\psi}{\di y_1}(I(\tilde u+v_\lambda+\xi e),\vert E\vert,\mu(\tilde
u+v_\lambda+\xi e,E))\notag\\
\times\eps_{km}(\tilde u+v_\lambda+\xi e)\eps_{km}(e)\eps_{ij}(\tilde u+v_
\lambda+\xi e)\eps_{ij}(e)]\,dx.  \label{5.48}
\end{gather}
By using \eqref{4.15} it is easy to see that
\begin{equation}\label{5.49}
\gamma_2(\xi)\ge 0,
\end{equation}
and Lemma 3.1 implies
\begin{equation}\label{5.50}
\gamma_4(\xi)\ge\mu_1\|e\|_X^2, \quad \mu_1=\min (2a_1,2a_3).
\end{equation}
\eqref{5.45} yields
\begin{equation}\label{5.51}
\vert f_\xi\vert\le 4\,\frac{\vert e\vert}{\alpha\sqrt{n}+\vert\tilde u+v_
\lambda+\xi e\vert}.
\end{equation}
We denote
\begin{equation}\label{5.52}
b_0=\sup\Big \vert\frac{\di b}{\di y_2}(\vert E(x)\vert,y_2)\Big\vert,
\quad y_2\in [0,1],\quad x\in\Omega.
\end{equation}
By \eqref{5.44}, \eqref{5.51}, and \eqref{5.52} we obtain
\begin{equation}\label{5.53}
\vert\gamma_1(\xi)\vert\le 8b_0\|e\|_X\Big(\int_\Omega\vert e\vert^4\,dx\Big)^
{\frac14}\Big(\int_\Omega(\alpha\sqrt{n}+\vert\tilde u+v_\lambda+\xi e\vert)
^{-4}\,\,dx\Big)^{\frac14}\le c_1 b_0\|e\| _X^2.
\end{equation}
Here
\begin{equation}\label{5.54}
c_1=8\check c\hat c,
\end{equation}
where $\check c$ is the constant of the inequality
\begin{equation}\label{5.55}
\|\vert e\vert\|_{L_4(\Omega)}\le\check c\|e\|_X, 
\end{equation}
and
\begin{equation}\label{5.56}
\hat c=\sup(\int_\Omega(\alpha\sqrt{n}+\vert\tilde u+v_\lambda+\xi e\vert)^
{-4}\,\,dx)^{\frac14}, \quad \xi\in (0,1),\quad \|e\|_X\le 2r,
\end{equation}
$r$ being the constant of \eqref{5.17}.
 
\eqref{5.42}, \eqref{5.47}, \eqref{5.51}, and \eqref{5.53} yield
\begin{equation}\label{5.57}
\vert\gamma_3(\xi)\vert\le 8\tilde c\|e\|_X\Big(\int_\Omega\vert e\vert^4\,\,dx
\Big)^{\frac14}\Big(\int_\Omega(\alpha\sqrt{n}+\vert\tilde u+v_\lambda+\xi e
\vert)^{-4}\,\,dx\Big)^{\frac14}\le c_1\tilde c\|e\|_X^2.
\end{equation}
Assume that
\begin{equation}\label{5.58}
\mu_1-c_1(b_0+\tilde c)=c_0>0.
\end{equation}
Then $\frac{d\eta}{dt}(\xi)\ge c_0\|e\|_X^2$, and by \eqref{5.41}, \eqref{5.43}
we obtain  that $e=0$. Thus, we proved the following:

\begin{th}
Suppose that the conditions $(C1)$ ($\phi$ replaced by $\psi)$, $(C4)$, 
\eqref{4.6}, \eqref{4.7}, \eqref{5.42}, \eqref{5.58} are satisfied. Then there
exists a unique solution of \eqref{5.5}--\eqref{5.7} in the ball
\begin{equation}\notag
d_r=\{u\in X, \quad \|u\|\le r=a_1^{-1}\|K+F\|_{X^*}\}.
\end{equation}
\end{th}
Note that in the case that the values of the function $\vert E\vert$ are small
the constants $b_0$ and $\tilde c$ are small, \eqref{5.58} is satisfied and
there exists a unique solution of \eqref{5.5}--\eqref{5.7}.

\section{Variational inequality for the extended Bingham electrorheological 
fluid.}

We now consider a problem on stationary flow of the extended Bingham 
electrorheological 
fluid. The constitutive equation of this fluid is the following:
\begin{equation}\label{6.1}
\phi(I(u),\vert E\vert,\mu(u,E))=\frac{b(\vert E\vert,\mu(u,E))}{I(u)^{\frac12}
} +b_1(\vert E\vert,\mu(u,E)).
\end{equation}
We deal with the problem \eqref{4.1}--\eqref{4.4} and assume that \eqref{4.5}
and \eqref{4.7} are satisfied. Then, according to Remark 4.1 the generalized 
solution of our problem is $u=\tilde u+v$, where $\tilde u$ is a function 
satisfying \eqref{4.6} and $v$ is a solution of the problem
\begin{gather}
v\in V,\label{6.2}\\
J(v,h)-J(v,v)+(L_1(v),h-v)\ge (K+F,h-v), \qquad  h\in V.\label{6.3}
\end{gather}
Here, $J$ is the functional given by \eqref{4.9} and the operator $L_1:X\to 
X^*$ is defined as follows
\begin{equation}\label{6.4}
(L_1(v),h)=2\int_\Omega b_1(\vert E\vert,\mu(\tilde u+v,E))\eps_{ij}(\tilde u+v)
\eps_{ij}(h)dx, \qquad v,h\in X.
\end{equation}
The function $b_1$ is subject to the following condition:
\begin{description}
\item[(C5)]
     $b_1: (y_1, y_2)\to b_1(y_1,y_2)$ is a continuous function  
     on $\R_+\times[0,1]$ and satisfies 
\begin{equation}\label{6.5}
a_6\le b_1(y_1,y_2)\le a_7, \qquad (y_1,y_2)\in\R_+\times[0,1],
\end{equation}
     with positive constants $a_6$ and $a_7$.
\end{description}
We approximate the functional $J$ by  $J_\lambda$ as given by \eqref{5.1}. 
Replacing $J$ by $J_\lambda$,  by analogy with the reasoning in the proof of 
Theorem 4.1 we obtain the following problem: find $v_\lambda$ such that
\begin{gather}
v_\lambda\in V, \label{6.6}\\
\Big(\frac{\di J_\lambda}{\di h}(v_\lambda,v_\lambda),w\Big)+(L_1(v_\lambda),
w)=(K+F,w), \qquad w\in V,\label{6.7}
\end{gather}
with $\frac{\di J_\lambda}{\di h}$ given by \eqref{5.2}.
\begin{th}
Suppose that conditions \eqref{4.6}, \eqref{4.7}, $(C4)$, $(C5)$ are satisfied.
Then there exists a solution $v$ of \eqref{6.2}, \eqref{6.3}. Moreover, for an
arbitrary $\lambda>0$ there exists a solution of \eqref{6.6}, \eqref{6.7}, and
there exists a function $p_\lambda$ such that the pair $(v_\lambda$, 
$p_\lambda)$ is a solution of the problem
\begin{gather}
(v_\lambda,p_\lambda)\in X\times L_2(\Omega), \label{6.8}\\
\Big(\frac{\di J_\lambda}{\di h}(v_\lambda,v_\lambda),w\Big)+(L_1(v_\lambda),w)
-(B^*p_\lambda,w)=(K+F,w), \qquad  w\in X,\label{6.9}\\
(Bv_\lambda,q)=0, \qquad  q\in L_2(\Omega).\label{6.10}
\end{gather}
A subsequence can be extracted from the sequence $\{v_\lambda\}$, again denoted
by $\{v_\lambda\}$, such that
\begin{equation}\label{6.11}
v_\lambda\rightharpoonup v\mbox{ in  } X \mbox{ and }v_\lambda\to
\mbox{ in } L_2(\Omega)^n \mbox{ as }\lambda\to 0.
\end{equation}
If $I(\tilde u+v)\ne 0$ almost everywhere in $\Omega$, then the functional 
$h\to J(v,h)$ is G\^ateaux differentiable at the point $v$, and there exists a
function $p\in L_2(\Omega)$ such that the pair $(v$, $p)$ is a solution of the 
problem 
\begin{gather}
v\in V, \qquad p\in L_2(\Omega),\label{6.12}\\
\Big(\frac{\di J}{\di h}(v,v),h\Big)+(L_1(v),h)-(B^*p,h)=(K+F,h),\qquad
 h\in X,\label{6.13}
\end{gather}
with $\frac{\di J}{\di h}$ given by \eqref{4.20}.
\end{th}
{\it Proof}. The existence of a solution $v_\lambda$ of \eqref{6.6}, 
\eqref{6.7} follows from Theorem 5.1 as well as the existence of a function 
$p_\lambda$ such that \eqref{6.8}--\eqref {6.10} hold. It is inferred from the
proof of Theorem 5.1 (see \eqref{5.16}) that
\begin{equation}\label{6.14}
v_\lambda \mbox{ remains in a bounded set of } V \mbox{ independent of } 
\lambda.
\end{equation}
Therefore, from sequence $\{v_\lambda\}$ we can select a subsequence, again
denoted by $\{v_\lambda\}$, such that
\begin{align}
&v_\lambda\rightharpoonup v \mbox{  in } X \mbox{ as } \lambda\to 0, \label{6.15}\\
&v_\lambda\to v \mbox{  in } L_2(\Omega)^n \mbox{ and a.e. in } \Omega.
\label{6.16}
\end{align}
For $h\in V$  we introduce
\begin{equation}\label{6.17}
Z_\lambda=(L_1(v_\lambda),h-v_\lambda)+J_\lambda(v_\lambda,h)-J_\lambda
(v_\lambda,v_\lambda)-(K+F,h-v_\lambda). 
\end{equation}
Using \eqref{6.9}, we see that
\begin{equation}\label{6.18}
Z_\lambda=-\Big(\frac{\di J_\lambda}{\di h}(v_\lambda,v_\lambda),h-v_\lambda
\Big)+J_\lambda(v_\lambda,h)-J_\lambda(v_\lambda,v_\lambda).
\end{equation}
It follows from \eqref{5.1}, \eqref{5.2} and Lemma 3.4 (cf. \eqref{3.23}) that 
for an arbitrarily fixed $w\in X$ the functional $u\to J_
\lambda(w,u)$ is convex. Therefore
\begin{equation}\label{6.19}
Z_\lambda\ge 0.
\end{equation}
(C5), \eqref{6.16} and the Lebesgue theorem give
\begin{equation}\label{6.20}
b_1(\vert E\vert,\mu(\tilde u+v_\lambda,E))\eps_{ij}(h)\to b_1(\vert E\vert,\mu
(\tilde u+v),E))\eps_{ij}(h)\quad\mbox{in } L_2(\Omega).
\end{equation}
\eqref{6.15}, \eqref{6.20} imply
\begin{equation}\label{6.21}
\lim(L_1(v_\lambda),h)=(L_1(v),h).
\end{equation}
We have
\begin{equation}\label{6.22}
(L_1(v_\lambda),v_\lambda)=A_{1\lambda}+A_{2\lambda},
\end{equation}
where
\begin{align}
&A_{1\lambda}=2\int_\Omega b_1(\vert E\vert,\mu(\tilde u+v_\lambda,E))\eps_{ij}
(\tilde u)\eps_{ij}(v_\lambda)dx, \label{6.23}\\
&A_{2\lambda}=2\int_\Omega b_1(\vert E\vert,\mu(\tilde u+v_\lambda,E))I
(v_\lambda)dx. \label{6.24}  
\end{align}
\eqref{6.20} still holds true if the function $h$ is replaced by $\tilde u$.
Consequently, \eqref{6.15} implies
\begin{equation}\label{6.25}
\lim A_{1\lambda}=2\int_\Omega b_1(\vert E\vert,\mu(\tilde u+v,E))\eps_{ij}
(\tilde u)\eps_{ij}(v).
\end{equation}
It follows from \eqref{6.5} and \eqref{6.16} that
\begin{equation}\notag
[b_1(\vert E\vert,\mu(\tilde u+v_\lambda, E))]^{\frac12} w\to[b_1(\vert E\vert,
\mu(\tilde u+v))] ^{\frac12} w \quad\mbox{in } L_2(\Omega), \quad w\in L_2(\Omega),
\end{equation}
so that \eqref{6.15} yields
\begin{gather}
\lim\int_\Omega[b_1(\vert E\vert,\mu(\tilde u+v_\lambda,E))]^{\frac12}\eps_{ij}
(v_\lambda)wdx=\int_\Omega[b_1(\vert E\vert,\mu(\tilde u+v,E))]^{\frac12}
\eps_{ij}(v)wdx, \notag\\
w\in L_2(\Omega).\notag
\end{gather}
Therefore,
\begin{equation}\label{6.26}
[b_1(\vert E\vert,\mu(\tilde u+v_\lambda,E))]^{\frac12}\eps_{ij}(v_\lambda)
\rightharpoonup [b_1(\vert E\vert,\mu(\tilde u+v,E))]^{\frac12}\eps_{ij}(v)
\quad \mbox{ in } L_2(\Omega).
\end{equation}
\eqref{6.24} and \eqref{6.26} give
\begin{equation}\label{6.27}
\lim\inf A_{2\lambda}\ge 2\int_\Omega b_1(\vert E\vert,\mu(\tilde u+v,E))I(v)dx.
\end{equation}
By \eqref{6.22}, \eqref{6.25}, and \eqref{6.27} we obtain
\begin{equation}\label{6.28}
\lim\inf (L_1(v_\lambda),v_\lambda)\ge(L_1(v),v).
\end{equation}
\eqref{5.1}, \eqref{6.16} and the Lebesgue theorem give
\begin{equation}\label{6.29}
\lim J_\lambda(v_\lambda,h)=J(v,h).
\end{equation}
Setting
\begin{gather}
b_\lambda=b(\vert E\vert,\mu(\tilde u+v_\lambda,E)),\qquad b_0=b(\vert E\vert,
\mu(\tilde u+v,E)),\label{6.30}\\
I_\lambda=I(\tilde u+v_\lambda), \qquad I_0=I(\tilde u+v),\label{6.31}
\end{gather}
we have
\begin{equation}\label{6.32}
J_\lambda(v_\lambda,v_\lambda)=J_\lambda(v,_\lambda)+B_{1\lambda},
\end{equation}
where
\begin{equation}\label{6.33}
B_{1\lambda}=2\int_\Omega (b_\lambda-b_0)(\lambda+I_\lambda)^
{\frac12}\,dx,
\end{equation}
and
\begin{equation}\label{6.34}
\vert B_{1\lambda}\vert\le 2\Big(\int_\Omega(\lambda+I_\lambda)
\,dx\Big)^{\frac12}\,\Big(\int_\Omega\vert b_\lambda-b_0\vert^2\,dx\Big)^
{\frac12}.
\end{equation}
\eqref{6.15}, \eqref{6.16} and \eqref{2.20} imply
\begin{equation}\label{6.35}
\lim B_{1\lambda}=0.
\end{equation}
\eqref{4.9} and \eqref{5.1} yield $J_\lambda(v,v_\lambda)\ge J(v,v_\lambda)$,
 whence
\begin{equation}\label{6.37}
\lim\inf J_\lambda(v,v_\lambda)\ge\lim\inf J(v,v_\lambda).
\end{equation}
\eqref{6.15} and Lemma 3.7 yield
\begin{equation}\label{6.38}
\lim\inf J(v,v_\lambda)\ge J(v,v).
\end{equation}
\eqref{6.32}, \eqref{6.35}, \eqref{6.37}, and \eqref{6.38} give
\begin{equation}\label{6.39}
\lim\inf J_\lambda(v_\lambda,v_\lambda)\ge J(v,v).
\end{equation}
\eqref{6.10}, \eqref{6.15} imply \eqref{6.2}, and using \eqref{6.17}, \eqref
{6.19}, \eqref{6.21}, \eqref{6.28}, \eqref{6.29}, \eqref{6.39} we obtain
\eqref{6.3}.

It follows from Remark 4.1 that if $I(\tilde u+v)\ne 0$ almost everywhere
in $\Omega$, then there exists a function $p$ such that \eqref{6.12}, \eqref
{6.13} hold.
$\blacksquare$

{\bf Remark 6.1}. Assume that in \eqref{6.3}, \eqref{6.7}, \eqref{6.9} the 
operator $L_1$ is replaced by the operator $L$ defined by \eqref{4.10}, and the
function $\psi$ meets one of the conditions (C1), (C2), (C3) with $\phi$ 
replaced by $\psi$. Then, by Theorem 5.1 for an arbitrary $\lambda>0$ 
there exists a solution ($v_\lambda$, $p_\lambda$) of \eqref{6.8}--\eqref{6.10}, 
and a subsequence $\{v_\lambda\}$ can be extracted satisfying \eqref{6.15}, 
\eqref{6.16}.

However, \eqref{6.15}, \eqref{6.16} do not imply $\lim\inf(L(v_\lambda),
v_\lambda)\ge (L(v),v)$ (compare with (\eqref{6.28}), and we cannot assert that 
$v$ is a solution of \eqref{4.12}, \eqref{4.13}. In the next section we
prove the existence of a solution of \eqref{4.12}, \eqref{4.13}
under conditions which are more restrictive than those of Theorem 5.1.

\section{General variational inequality.}
We assume that the function $\mu$ in the operator $L$ defined by \eqref{4.10} 
is replaced by a function $\mu_1$ such that
\begin{equation}\label{7.1}
u_m\rightharpoonup u \quad\mbox{in  } X\Rightarrow
\mu_1(u_m,E)\to\mu_1(u,E)\mbox{ in } L_\infty(\Omega).
\end{equation}
According to \eqref{2.7} we may define $\mu_1$ as follows:
\begin{equation}\label{7.2}
\mu_1(u,E)(x)=\Big(\frac{\alpha\tilde I+Pu(x)+\check u}{\alpha\sqrt n+\vert P
u(x)+\check u\vert},\,\,\frac{\beta\tilde I+PE(x)}{\beta\sqrt n+\vert PE(x)
\vert}\Big)_{\R^n}^2,
\end{equation}
where $\alpha$, $\beta$ are small positive constants, $\tilde I$ is a vector
with components equal to one, and $P$  an operator of regularization given by
\begin{equation}\label{7.3}
Pu(x)=\int_{\R^n} \omega(\vert x-x'\vert)u(x')dx',\qquad x\in\overline\Omega,
\end{equation}
where
\begin{align}
&\omega\in C^\infty(\R_+),\quad \mbox{supp }\omega=[0,a],\quad \omega(z)\ge 0 \quad
z\in\R_+, \notag\\
&\int_{\R^n}\omega(\vert x\vert)dx=1,\quad a \mbox{ is a small positive
constant}.\label{7.4} 
\end{align}
In \eqref{7.3} we assume that the function $u$ is extended to $\R^n$.
In case that $Pu(x)\ne 0$ a.e. in $\Omega$ we may choose $\alpha=0$, if
$PE(x)\ne 0$ a.e. in $\Omega$ we may choose $\beta=0$ .

For the function $\mu_1$ condition \eqref{7.1} is
satisfied. From the physical point of view \eqref{7.2} means that the value
of the function $\mu_1$ and therefore the viscosity of the fluid at a point
$x$ depends on the angle between the vectors of velocity and electric field
strength at points belonging to some small  vicinity of the point $x$, implying
that the  model is not local.

This seems to be natural, since electrorheological
properties of the fluid are linked with the presence of small solid particles
in the fluid. The mean dimension of these particles can be taken as the
regularization parameter $a$.

In the case under consideration the operator $L$ is defined as follows:
\begin{equation}\label{7.5}
(L(v),h)=2\int_\Omega \psi(I(\tilde u+v),\vert E\vert,\mu_1(\tilde u+v,E))
\eps_{ij}(\tilde u+v)\eps_{ij}(h)dx, \mbox{ } v,h\in X.
\end{equation}
We assume also that the following condition of uniform continuity of the
function $\psi$ holds:
\begin{description}
\item[(C6)]
          for an arbitrary $\gamma>0$  there exists  $\eps>0$
          such that the conditions  
\begin{equation}\notag
y'_3, y''_3\in [0,1], \quad\vert y'_3-y''_3\vert\le\eps,\quad y_1,y_2\in\R_+ 
\end{equation}
imply 
\begin{equation}\notag
  \vert\psi(y_1,y_2,y'_3)-\psi(y_1,y_2,y''_3\vert\le\gamma.
\end{equation}
\end{description}

The function $\mu$ can as well be replaced by the function $\mu_1$ in the 
functionals $J$ and $J_\lambda$. In the following theorem this is not assumed,
although it is also valid in this case.

\begin{th} Suppose that the conditions $(C4)$, \eqref{4.6}, \eqref{4.7} 
are satisfied and assume that the function $\psi$ meets one of the conditions $(C1)$, 
$(C2)$, $(C3)$  ($\phi$ replaced by $\psi$) and that $(C6)$ holds. Further
assume that the function $\mu_1$ meets condition \eqref{7.1}, and the operator
$L$ is given by \eqref{7.5}. Then, for an arbitrary $\lambda>0$ there
exists a solution $(v_\lambda$, $p_\lambda)$ of \eqref{5.5}--\eqref{5.7}, and 
there exists a solution $v$ of \eqref{4.12}, \eqref{4.13}. A subsequence can be
selected from the sequence $\{v_\lambda\}$, again denoted by $\{v_\lambda\}$, 
such that
\begin{equation}\notag
v_\lambda\rightharpoonup v \mbox{ in } X \mbox{ and } v_\lambda\to v 
\mbox{ in } L_2(\Omega)^n\mbox{ as }\lambda\to 0.
\end{equation}
\end{th}
{\it Proof}. 1) The existence of a solution ($v_\lambda$, $p_\lambda$) of 
\eqref{5.5}--\eqref{5.7} follows from Theorem 5.1, and it is 
inferred from the proof of this theorem (see \eqref{5.16}), that
$v_\lambda$ remains in a bounded set of $V$ independent of $\lambda$. Therefore,
from the sequence $\{v_\lambda\}$ we can select a subsequence, again denoted
by $\{v_\lambda\}$, such that
\begin{align}
&v_\lambda\rightharpoonup v \mbox{ in } V \mbox{ as } \lambda\to 0,\label{7.6}\\
&v_\lambda\to v \mbox{  in } L_2(\Omega)^n \mbox{ and a.e. in } \Omega.
\label{7.7}
\end{align}
For every $v_\lambda$ we define a functional $\Psi_\lambda$ as follows:
\begin{equation}\label{7.8}
\Psi_\lambda(v)=J_\lambda(v_\lambda,v)+\Phi(v_\lambda,v)-(K+F,v),\quad v\in V,
\end{equation}
where
\begin{equation}\label{7.9}
\Phi(v_\lambda,v)=\int_\Omega\Big(\int_0^{I(\tilde u+v)}\psi(\xi,\vert E\vert,
\mu_1(\tilde u+v_\lambda,E))d\xi\Big)dx.
\end{equation}
Consider the problem: find a function $\tilde v$ satisfying
\begin{equation}\label{7.10}
\tilde v\in V,\quad \Psi_\lambda(\tilde v)=\min_{v\in V}\,\Psi_\lambda(v).
\end{equation}
If $\tilde v$ is a solution of \eqref{7.10}, then we have
\begin{equation}\label{7.11}
\tilde v\in V, \quad \frac{d}{dt}\Psi_\lambda(\tilde v+th)\Big\vert_{t=0}\,=
\Big(\frac{\di J_\lambda}{\di v}(v_\lambda,\tilde v),h)+(L(v_\lambda,\tilde v),
h\Big)-(K+F,h)=0,\quad h\in V.
\end{equation}
Here, $L(v_\lambda,\tilde v)$ is the G\^ateaux derivative of the functional 
$\Phi(v_\lambda,.):u\to\Phi(v_\lambda,u)$  at the point $\tilde v$, i.e.
\begin{equation}\notag
\frac{\di \Phi}{\di v}(v_\lambda,\tilde v)=L(v_\lambda,\tilde v).
\end{equation} 
The operator $L(v_\lambda,.):X\ni u\to L(v_\lambda,u)\in X^*$ has the form
\begin{equation}\label{7.12}
(L(v_\lambda,u),h)=2\int_\Omega\psi(I(\tilde u+u),\vert E\vert,\mu_1(\tilde u
+v_\lambda,E))\eps_{ij}(\tilde u+u)\eps_{ij}(h)dx, \quad u,h\in X,
\end{equation}
and \eqref{7.5} yields
\begin{equation}\label{7.13}
L(v,v)=L(v).
\end{equation}
It follows from \eqref{5.3}, \eqref{5.4} that the function $\tilde v=v_\lambda$
is a solution of \eqref{7.11}. By means of Lemmas 3.1--3.4 the
functional $\Psi_\lambda$ is strictly convex. Therefore, there exists a unique 
solution $\tilde v=v_\lambda$ of \eqref{7.11},  and problems \eqref{7.10} and 
\eqref{7.11} are equivalent.

\eqref{7.10} implies
\begin{equation}\label{7.14}
J_\lambda(v_\lambda,h)+\Phi(v_\lambda,h)-J_\lambda(v_\lambda,v_\lambda)-\Phi
(v_\lambda,v_\lambda)\ge(K+F,h-v_\lambda),\quad  h\in V.
\end{equation}
2) It follows from \eqref{7.6}, \eqref{7.7} and the proof of Theorem 6.1 (see
\eqref{6.29}, \eqref{6.39}) that
\begin{equation}\label{7.15}
\lim J_\lambda(v_\lambda,h)=J(v,h),\qquad
\lim\inf J_\lambda(v_\lambda,v_\lambda)\ge J(v,v).
\end{equation}
We have
\begin{equation}\label{7.16}
\Phi(v_\lambda,h)=\int_\Omega f_\lambda dx, \quad \Phi(v,h)=\int_\Omega f\,dx,
\end{equation}
where
\begin{align}
f_\lambda(x)=&\int_0^{I(\tilde u+h)(x)}\,\psi(\xi,\vert E(x)\vert,\mu_1
(\tilde u+v_\lambda,E)(x))d\xi, \notag\\
f(x)=        &\int_0^{I(\tilde u+h)(x)}\,\psi(\xi,\vert E(x)\vert,\mu_1
(\tilde u+v,E)(x))d\xi.\notag
\end{align}
\eqref{7.7} and (C6) imply $f_\lambda\to f$ almost everywhere in $\Omega$, and
\eqref{2.16} yields
\begin{equation}\notag
\vert f_\lambda\vert\le a_2 I(\tilde u+h).
\end{equation}
Thus \eqref{7.16} and the Lebesgue theorem give
\begin{equation}\label{7.17}
\lim\Phi(v_\lambda,h)=\Phi(v,h).
\end{equation}
It is obvious that
\begin{align}
\Phi(v_\lambda,v_\lambda)=\Phi(v,_\lambda)+\alpha_\lambda,\label{7.18}\\
\alpha_\lambda=\Phi(v_\lambda,v_\lambda)-\Phi(v,_\lambda)=\int_\Omega\Big\{
\int_0^{I(\tilde u+v_\lambda)}[&\psi(\xi,\vert E\vert,(\mu_1(\tilde u+v_\lambda,
E))\notag\\
-&\psi(\xi,\vert E\vert,\mu_1(\tilde u+v,E))]d\xi\Big\}dx. \notag
\end{align}
\eqref{7.1}, \eqref{7.6} and (C6) imply 
\begin{equation}\notag
\alpha_\lambda\le\beta_\lambda\int_\Omega I(\tilde u+v_\lambda)dx,
\end{equation}
and $\lim\beta_\lambda=0$. Therefore $\lim\alpha_\lambda=0$.

The functional $\Phi(v,.):u\to\Phi(v,u)$ is continuous in $X$. Indeed, let 
$u_m\to u$ in $X$. We have
\begin{gather}
\Big\vert\Phi(v,u_m)-\Phi(v,u)\Big\vert=\Big\vert\int_\Omega\Big(\int_{I(\tilde
u+u)}^{I(\tilde u+u_m)}\,\psi(\xi,\vert E\vert,\mu_1(\tilde u+v,E))d\xi\Big)
dx\Big\vert \notag\\
\le a_2\Big\vert\int_\Omega(I(\tilde u+u_m)-I(\tilde u+u))dx\Big
\vert,\label{7.19}
\end{gather}
and the right hand side of this inequality tends to zero as $m\to\infty$.

By Lemmas 3.1, 3.2, 3.3 the functional $\Phi(v,.)$ is convex in $X$. Therefore,
$\Phi(v,.)$ is lower semi-continuous for the weak topology on $X$, and \eqref
{7.6}, \eqref{7.18} imply
\begin{equation}\label{7.20}
\lim\inf\Phi(v_\lambda,v_\lambda)\ge\Phi(v,v).
\end{equation}
By \eqref{7.15}, \eqref{7.17}, \eqref{7.20} we pass to the limit  as $\lambda
\to 0$ in \eqref{7.14}. This gives
\begin{equation}\label{7.21}
J(v,h)+\Phi(v,h)-J(v,v)-\Phi(v,v)\ge(K+F,h-v), \quad  h\in V.
\end{equation}
Taking into account \eqref{7.21} and the convexity of the functional $J(v,.):
u\to J(v,u)$, we get
\begin{gather}
J(v,v)+\Phi(v,v)-(K+F,v)   \notag\\
\le J(v,(1-\theta)v+\theta h)+\Phi(v,(1-\theta)v+\theta 
h)-(K+F,(1-\theta)v+\theta h)  \notag\\
\le(1-\theta)J(v,v)+\theta J(v,h)+\Phi(v,(1-\theta)v+\theta h)-(K+F,(1-\theta)
v+\theta h), \theta\in (0,1),\notag
\end{gather} 
whence
\begin{equation}\label{7.22}
\frac{\Phi(v,(1-\theta)v+\theta h)-\Phi(v,v)}{\theta}+J(v,h)-J(v,v)-(K+F,h-v)
\ge 0.
\end{equation}
For $\theta\to 0$ we get \eqref{4.13}, with the operator $L$ defined by
\eqref{7.12}, \eqref{7.13}.
$\blacksquare$

{\bf Remark 7.1}. We have shown that \eqref{7.21} implies \eqref{4.13}. 
Let us show that \eqref{4.13} yields \eqref{7.21}, that is problems 
\eqref{4.12}, \eqref{4.13} and \eqref{4.12}, \eqref{7.21} are equivalent.

Let \eqref{4.12}, \eqref{4.13} be valid. Obviously,
\begin{gather}
J(v,h)+\Phi(v,h)-J(v,v)-\Phi(v,v)-(K+F,h-v) \notag\\
=J(v,h)-J(v,v)+(L(v,v),h-v)+\Phi(v,h)-\Phi(v,v)\notag\\
-(L(v,v),h-v)-(K+F,h-v) \label{7.23}
\end{gather}
The functional $u\to\Phi(v,u)$ is convex, whence
\begin{equation}\label{7.24}
\Phi(v,h)-\Phi(v,v)-(L(v,v),h-v)\ge 0.
\end{equation}
\eqref{4.13}, \eqref{7.13}, \eqref{7.23}, \eqref{7.24} give \eqref{7.21}.

{\bf Remark 7.2}. By comparing Theorems 5.1 and 7.1 we observe that a solution
of the operator equations for a fluid with  bounded viscosity function
\eqref{2.12a} exists under less restrictive conditions than the conditions for
the existence of a solution of the variational inequality \eqref{4.13} for a
fluid with  unbounded viscosity function \eqref{2.12}. In addition, such an
important characteristic of the flow as the function of pressure is defined for
 \eqref{2.12} only in case that \eqref{4.17} holds, i.e., when the
viscosity function d\oe s not take infinite values.
But in this case, the variational inequality reduces to operator equations
as outlined in  Section 4.

Moreover, from a physical point of view a fluid with finite viscosity 
\eqref{2.12a} seems to be more reasonable than a fluid with  unbounded
viscosity \eqref{2.12}.

\section{Problems with given function $\mu$.}
In the case that the distance between the electrodes is small compared with the 
lengths of the electrodes, one can assume that in between the electrodes the 
velocity vector is orthogonal to the vector of electric field strength, and the 
electric fields strength is equal to zero in the remaining  part of the domain
under consideration.

In this case one reckon that $\mu(u,E)$ is a known function of $x$, so that the
viscosity functions \eqref{2.12} and \eqref{2.12a} take the form
\begin{gather}
\phi(I(u),\vert E\vert,x)=\frac{e(\vert E\vert,x)}{I(u)^{\frac12}}+\psi_1(I(u),
\vert E\vert,x),  \label{8.1}\\
\phi(I(u),\vert E\vert,x)=e(\vert E\vert,x)(\lambda+I(u))^{-\frac12}\,+\psi_1
(I(u),\vert E\vert,x), \label{8.2}
\end{gather}
and the constitutive equation is defined by \eqref{2.1}.

Dependence of the viscosity function on $x$ in \eqref{8.1}, \eqref{8.2} is connected with the anisotropy of the fluid. If the direction of the velocity vector at each point $x$ at which $E(x)\ne 0$ is known, then the viscosity functions
 \eqref{2.12} and \eqref{2.12a} transform in relations \eqref{8.1}, \eqref{8.2}.

We assume the function $\psi_1$ to satisfy 

\begin{description}
\item[(CO)]
         for almost all $x\in\Omega$ the function $\psi_1(.,.,x):(y_1,y_2)\to
        \psi_1(y_1,y_2,x)$ is continuous in $\R_+^2$, and for an arbitrarily 
         fixed          $(y_1,y_2) \in\R_+^2$ the function 
         $\psi_1(y_1,y_2,.):x\to\psi_1(y_1, 
         y_2,x)$ is measurable in $\Omega$. \\

         We also suppose that for almost all
         $x\in\Omega$ 
         and all $y_2\in\R_+$, the function $\psi_1(.,y_2,x):y_1\to\psi_1(y_1,y_2,
         x)$ satisfies one of the following conditions (C1a), (C2a), (C3a):
\end{description}

\begin{description}
\item[(C1a)]
         $\psi_1(.,y_2,x)$ is continuously differentiable in $\R_+$ and the 
         following inequalities hold:
\begin{gather}
a_2\ge\psi_1(y_1,y_2,x)\ge a_1, \label{8.3}\\ 
\psi_1(y_1,y_2,x)+2\frac{\di\psi_1}{\di y_1}(y_1,y_2,x)\ge a_3 \label{8.4}\\
\Big\vert\frac{\di\psi_1}{\di y_1}(y_1,y_2,x)\Big\vert y_1\le a_4.\label{8.5}
\end{gather}
\end{description}

\begin{description}
\item[(C2a)]
         \eqref{8.3} is fulfilled and for an arbitrary $(z_1,z_2)\in\R_+^2$
          the following inequality is valid:
\begin{equation}\label{8.6}
[\psi_1(z_1^2,y_2,x)z_1-\psi_1(z_2^2,y_2,x)z_2](z_1-z_2)\ge a_3(z_1-z_2)^2.
\end{equation}
\end{description}

\begin{description}
\item[(C3a)]
         \eqref{8.3} is fulfilled and the function $z\to\psi_1(z^2,y_2,x)z$
          is strictly increasing in $\R_+$, i.e., the conditions $z_1,z_2\in
          \R_+$, $z_1>z_2$ imply $\psi_1(z_1^2,y_2,x)z_1>\psi_1(z_2^2,y_2,
          x)z_2$.
\end{description}

(C1a), (C2a), (C3a) are analogies of conditions (C1), (C2), (C3), and an 
analog of (C4) is the following condition:
\begin{description}
\item[(C4a)]
           for almost all $x\in\Omega$ the function $e(.,x):y\to e(y,x)$  is
           continuous in $\R_+$ and for an arbitrarily fixed $y\in\R_+$, the 
           function $e(y,.):x\to e(y,x)$ is measurable in $\Omega$ and
\begin{equation}\label{8.7}
0\le e(y,x)\le a_5.
\end{equation}
\end{description}
Define functionals $Y$ and $Y_\lambda$, $\lambda>0$, as follows:
\begin{gather}
Y(u)=2\int_\Omega e(\vert E\vert,x)I(\tilde u+u)^{\frac12}\,dx, \quad u\in X,
\label{8.8}\\
Y_\lambda(u)=2\int_\Omega e(\vert E\vert,x)(\lambda+I(\tilde u+u))^{\frac12}\,
dx, \quad u\in X. \label{8.9}
\end{gather}
Define also an operator $L_2:X\to X^*$ by means of
\begin{equation}\label{8.10}
(L_2(u),h)=2\int_\Omega\psi_1(I(\tilde u+u),\vert E\vert,x)\eps_{ij}(\tilde u+
u)\eps_{ij}(h)dx, \quad u,h\in X. 
\end{equation}
Consider the following two problems:

Problem 1. Find a pair of functions $(v_\lambda$, $p_\lambda)$ such that
\begin{gather}
v_\lambda\in X, \quad p_\lambda\in L_2(\Omega), \label{8.11}\\
\Big(\frac{\di Y_\lambda}{\di u}(v_\lambda),h\Big)+(L_2(v_\lambda),h)-(B^*
p_\lambda,h)=(K+F,h), \quad  h\in X, \label{8.12}\\
(Bv_\lambda,q)=0, \qquad  q\in L_2(\Omega). \label{8.13}
\end{gather}

Problem 2. Find a function $v$ such that
\begin{gather}
v\in V,  \label{8.14}\\
Y(h)-Y(v)+(L_2(v),h-v)\ge(K+F,h-v), \quad  h\in V.\label{8.15}
\end{gather}
Here, the operator $\frac{\di Y_\lambda}{\di u}:X\to X^*$ is given by 
\begin{equation}\label{8.15a}
\Big(\frac{\di Y_\lambda}{\di u}(u),h\Big)=2\int_\Omega e(\vert E\vert,x)
(\lambda+I(\tilde u+u))^{-\frac12}\eps_{ij}(\tilde u+u)\eps_{ij}(h)dx, \qquad
u,h\in X.
\end{equation}
If $(v_\lambda$, $p_\lambda)$ is a solution of Problem 1, then the pair
($\tilde u+v_\lambda$, $p_\lambda$) is a generalized solution of 
\eqref{4.1}--\eqref{4.4} with the viscosity function defined by \eqref{8.2}.
If $v$ is a solution of Problem 2, then $\tilde u+v$ is a generalized
solution of \eqref{4.1}--\eqref{4.4} with the viscosity function 
defined by \eqref{8.1}.

\begin{th}
Suppose that the conditions \eqref{4.6}, \eqref{4.7}, $(C4a)$ are satisfied, and let
the function $\psi_1$ satisfy both $(C0)$ and one of the conditions $(C1a)$, 
$(C2a)$, $(C3a)$.
Then, for an arbitrary $\lambda>0$ there exists a unique solution $(v_\lambda$, 
$p_\lambda)$ of \eqref{8.11}--\eqref{8.13}. Moreover, there exists a 
unique solution $v$ of  \eqref{8.14}--\eqref{8.15}. In addition 
$v_\lambda\rightharpoonup v$ in $V$ as $\lambda\to 0$.
\end{th}
{\it Proof}. The existence of a solution ($v_\lambda$, $p_\lambda$) of 
\eqref{8.11}--\eqref{8.13} for an arbitrary $\lambda>0$ follows from
Theorem 5.1. Let ($v_\lambda^1$, $p_\lambda^1$) and 
($v_\lambda^2$, $p_\lambda^2$) be two solutions of \eqref{8.11}--\eqref{8.13}.
By \eqref{8.12} we obtain
\begin{equation}\label{8.16}
\Big(\frac{\di Y_\lambda}{\di u}(v_\lambda^1)+L_2(v_\lambda^1)-\frac{\di Y_
\lambda}{\di u}(v_\lambda^2)-L_2(v_\lambda^2),v_\lambda^1-v_\lambda^2\Big)=0.
\end{equation}
By Lemmas 3.1--3.4, the operator $\frac{\di Y_\lambda}{\di u}+L_2$ is
strictly monotone . Consequently \eqref{8.16} implies $v_\lambda^1=v_\lambda^2$,
whence $p_\lambda^1=p_\lambda^2$.

By Theorem 7.1 from the sequence $\{v_\lambda\}$ a subsequence, again 
denoted by $\{v_\lambda\}$, can be selected such that $v_\lambda\rightharpoonup
v$ in $X$ where $v$ is a solution of \eqref{8.14}, \eqref{8.15}. Also, 
Remark 7.1 infers
\begin{equation}\label{8.17}
v\in V,\quad Y(h)+\Phi_1(h)-Y(v)-\Phi_1(v)\ge(K+F,h-v), \quad h\in V,
\end{equation}
where
\begin{equation}\label{8.18}
\Phi_1(u)=\int_\Omega\Big(\int_0^{I(\tilde u+u)}\psi_1(\xi,\vert E\vert,x)\,
d\xi\Big)dx, \quad u\in V,
\end{equation}
and
\begin{equation}\label{8.19}
\Big(\frac{\di\Phi_1}{\di u}(u),h\Big)=(L_2(u),h).
\end{equation}
In addition, the problems \eqref{8.14}, \eqref{8.15} and  \eqref{8.17} are
equivalent. The functional $Y$ is convex, and the functional $\Phi_1$ is 
strictly convex. Therefore the functional
\begin{equation}\notag
\Psi_1(u)=Y(u)+\Phi_1(u)-(K+F,u), \qquad u\in V,
\end{equation}
is strictly convex, and if $v_1$, $v_2$ are two solutions of the problem 
\eqref{8.17}, then we have
\begin{equation}\notag
\Psi_1(\frac12(v_1+v_2))<\frac12\Psi_1(v_1)+\frac12\Psi_1(v_2)=\inf_{h\in V}
\Psi_1(h).
\end{equation}
Whence $v_1=v_2$.
$\blacksquare$

\section{Numerical solution of stationary problems.}
\subsection{Theorem on convergence.}

Let $\{X_m\}$, $\{N_m\}$ be sequences of finite-dimensional subspaces in $X$
and $L_2(\Omega)$, respectively, such that \eqref{3.38}, \eqref{3.39}, 
\eqref{3.43} and \eqref{5.11} hold. We denote
\begin{equation}\label{9.1}
L_3=\frac{\di Y_\lambda}{\di u}+L_2,
\end{equation}
where $L_2$ and $\frac{\di Y_\lambda}{\di u}$ are the operators defined by
\eqref{8.10}, \eqref{8.15a}.

We seek an approximate solution ($v_m$, $p_m$) of problem \eqref{8.11}--\eqref
{8.13} of the form
\begin{gather}
(v_m,p_m)\in X_m\times N_m, \label{9.2}\\
(L_3(v_m),h)-(B_m^*\,p_m,h)=(K+F,h),\quad h\in X_m, \label{9.3}\\
(B_m\,v_m,q)=0, \quad q\in N_m. \label{9.4}
\end{gather}
We remind that the operator $B_m$ is defined by \eqref{3.40} and $B_m^*$ is the
adjoint operator of $B_m$.
\begin{th}
Suppose that conditions \eqref{4.6}, \eqref{4.7}, $(C4a)$ are satisfied, and 
let the function $\psi_1$ satisfy both $(C0)$ and $(C3a)$. Let also \eqref{3.38},
\eqref{3.39}, \eqref{3.43} and \eqref{5.11} are fulfilled.
Then, for an arbitrary $m\in \N$ there exists a unique solution of \eqref{9.2}
--\eqref{9.4} and
\begin{equation}\label{9.5}
v_m\rightharpoonup v_\lambda \mbox{ in } X, \qquad
p_m\rightharpoonup p_\lambda \mbox{ in } L_2(\Omega), 
\end{equation}
where $v_\lambda$, $p_\lambda$ is the solution of \eqref{8.11}--\eqref{8.13}. 
If in addition $\psi_1$ satisfies $(C1a)$ or $(C2a)$, then
\begin{align}
&v_m\to v_\lambda \mbox{ in } X, \label{9.6} \\
&p_m\to p_\lambda \mbox{ in } L_2(\Omega). \label{9.7}
\end{align}
\end{th}
{\it Proof.} The existence of a unique solution $(v_m,p_m)$ of the problem 
\eqref{9.2}--\eqref{9.4} and the relations \eqref{9.5}  follows
from Theorems 5.1, 8.1. The following equalities also arise from the proof of
Theorem 5.1 (see \eqref{5.36}, \eqref{5.37})
\begin{gather}
(L_3(v_m),v_m)=(K+F,v_m)\to (K+F,v_\lambda), \label{9.8}\\
\lim(L_3(v_m),h)-(B^*\,p_\lambda,h)=(K+F,h),\quad h\in X. \label{9.9}
\end{gather}
\eqref{8.12}, \eqref{9.8}, \eqref{9.9} yield
\begin{equation}\label{9.10}
\lim(L_3(v_m)-L(v_\lambda),v_m-v_\lambda)=0.
\end{equation}
Assume that $(C2a)$ is satisfied; if $\psi_1$ meets $(C1a)$ it meets also
$(C2a)$ (see Subsection 2.2). Then observing Lemmas 3.2, 3.4, we obtain
\begin{equation}\notag
(L_3(v_m)-L_3(v_\lambda),v_m-v_\lambda)\ge 2a_3\int_\Omega [I(\tilde u+v_m)^
{\frac12}-I(\tilde u+v_\lambda)^{\frac12}]^2\,dx,
\end{equation}
and \eqref{9.10} implies
\begin{equation}\label{9.11}
\int_\Omega[I(\tilde u+v_m)^{\frac12}-I(\tilde u+v_\lambda)^{\frac12}]^2\,dx
\to 0.
\end{equation}
From here, taking into account that the function $u\to\int_\Omega u^2\,dx$ is 
a continuous mapping from $L_2(\Omega$ into $\R$, we obtain
\begin{equation}\label{9.12}
\int_\Omega I(\tilde u+v_m)dx\to \int_\Omega I(\tilde u+v_\lambda)dx.
\end{equation}
It is obvious that
\begin{gather}
\|v_m-v_\lambda\|_X^2=\int_\Omega\sum_{i,j=1}^n\,(\eps_{ij}(v_m-v_\lambda))^2\,
dx=\int_\Omega\sum_{i,j=1}^n[\eps_{ij}(\tilde u+v_m)-\eps_{ij}(\tilde u+v_
\lambda)]^2\,dx  \notag\\
=\int_\Omega I(\tilde u+v_m)dx+\int_\Omega I(\tilde u+v_\lambda)dx-2\int_
\Omega\eps_{ij}(\tilde u+v_m)\,\eps_{ij}(\tilde u+v_\lambda)dx. \label{9.13}
\end{gather}
By virtue of \eqref{9.5}, \eqref{9.12} the right-hand side of \eqref{9.13} 
tends to zero. Therefore, \eqref{9.6} holds true.

It follows from \eqref{8.12} and \eqref{9.1} that
\begin{equation}\label{9.14}
(L_3(v_\lambda),h)-(B^*\,p_\lambda,h)=(K+F,h), \quad h\in X_m,
\end{equation}
By \eqref{9.3} and \eqref{9.14} we get
\begin{equation}\notag
(B^*(p_m-\mu),h)=(L_3(v_m)-L_3(v_\lambda),h)+(B^*(p_\lambda-\mu),h), \quad
h\in X_m, \quad \mu\in N_m.
\end{equation}
This equality, together with \eqref{3.43}, yields
\begin{gather}
\|p_m-\mu\|_{L_2(\Omega)}\le\sup_{h\in X_m}\frac{(B^*(p_m-\mu),h)}
{\beta\|h\|_X} \notag\\
\le\beta^{-1}(\|L_3(v_m)-L_3(v_\lambda)\|_{X^*}\,+c\|p_\lambda-\mu\|_{L_2
(\Omega)}), \quad \mu\in N_m, \label{9.15}
\end{gather}
where
\begin{equation}\notag
c=\|B^*\|_{\cal L(L_2(\Omega),X^*)}=\|B\|_{\cal L(X,L_2(\Omega))}.
\end{equation}
Hence
\begin{gather}
\|p_\lambda-p_m\|_{L_2(\Omega)}\le\|p_\lambda-\mu\|_{L_2(\Omega)}+\|p_m-\mu\|
_{L_2(\Omega)} \notag\\
\le\beta^{-1}\|L_3(v_m)-L_3(v_\lambda)\|_{X^*}\,+(c\beta^{-1}+1)\inf_{\mu\in N_m}
\|p_\lambda-\mu\|_{L_2(\Omega)}.  \label{9.16}
\end{gather}
Lemmas 3.2, 3.4 and \eqref{9.6} imply $L_3(v_m)\to L_3(v_\lambda)$ in $X^*$,
and \eqref{9.7} follows from \eqref{9.16} and \eqref{3.39}.

\subsection{A saddle-point approach.}

We introduce two functionals 

$J:X\to\R$, $\,\,\Psi:X\times L_2(\Omega)\to\R$ defined by
\begin{gather}
J(u)=\int_\Omega\Big(\int_0^{I(\tilde u+u)}[e(\vert E\vert,x)(\lambda+\xi)^{-
\frac12}\,+\psi_1(\xi,\vert E\vert,x)]d\xi\Big)dx-(K+F,u), \label{9.17}\\
\Psi(u,\nu)=J(u)-(B^*\nu,u).  \label{9.18}
\end{gather}
Problem: find a saddle-point of the Lagrangian $\Psi$, i.e.
\begin{gather}
v_\lambda,p_\lambda\in X\times L_2(\Omega),  \label{9.19}\\
\Psi(v_\lambda,\nu)\le\Psi(v_\lambda,p_\lambda)\le\Psi(u,p_\lambda), \quad u\in 
X, \quad \mu\in L_2(\Omega).  \label{9.20}
\end{gather}
\begin{th}
Suppose that conditions \eqref{4.6}, \eqref{4.7}, $(C4a)$ are satisfied, and 
let the function $\psi_1$ meet $(C0)$ and one of the conditions $(C1a)$, $(C2a)$,
$(C3a)$. Then, for an arbitrary $\lambda>0$ there exists a unique solution 
$v_\lambda$, $p_\lambda$ of problem \eqref{9.19}, \eqref{9.20} which is the
solution of problem \eqref{8.11}--\eqref{8.13}, and the problems \eqref{8.11}--
\eqref{8.13} and \eqref{9.19}, \eqref{9.20} are equivalent.
\end{th}
{\it Proof.} Let $h$ be an arbitrary fixed element of $X$. Consider the function
$g:t\to g(t)=\Psi(v_\lambda+th,p_\lambda)$, $t\in\R$. It follows from the second
inequality in \eqref{9.20} that $g(0)\le g(t)$, $t\in\R$. The function $g$ is
differentiable in $\R$, and so $\frac{dg}{dt}(0)=0$. This equality is equivalent
to \eqref{8.12}.By Lemmas 3.1--3.4 the operator $L_3$ is monotone, and so the
functional $u\to\Psi(u,q)$ is convex for an arbitrarily fixed $q\in L_2(\Omega)$
(see e.g.\cite{13}). Therefore, the minimum of the functional $u\to\Psi(u,p_
\lambda)$  is characterized by \eqref{8.12}.

The first inequality in \eqref{9.20} gives $(Bv_\lambda,p_\lambda-\nu)\le 0$,
for all $\nu\in L_2(\Omega)$, and so we get \eqref{8.13}. However in this case
$\Psi(v_\lambda,p_\lambda)=\Psi(v_\lambda,\nu)$ for all $\nu\in L_2(\Omega)$.
Thus, the problems \eqref{8.11}--\eqref{8.13} and \eqref{9.19}, \eqref{9.20}
are equivalent. The existence and uniqueness of the solution of \eqref{9.19},
\eqref{9.20} follows from Theorem 8.1.
$\blacksquare$

Now we introduce the augmented Lagrangian as follows:
\begin{equation}\label{9.21}
\Psi_1(u,\nu)=\Psi(u,\nu)+\frac r2(Bu,Bu), \quad (u,\nu)\in X\times L_2(\Omega).
\end{equation}
It is obvious that the pair $(v_\lambda,p_\lambda)$ is also the saddle point of
the functional $\Psi_1$, where $r$ is an arbitrary positive constant.

We study

Algorithm of the augmented Lagrangian: find a sequence $\{v_m,p_m\}$ satisfying
\begin{gather}
(v_{m+1},p_{m+1})\in X\times L_2(\Omega), \label{9.22}\\
(L_3(v_{m+1}),h)-(B^*\,p_m,h)+r(Bv_{m+1},Bh)=(K+F,h), \quad h\in X, 
\label{9.23}\\
(p_{m+1}-p_m,\nu)+\rho_m(Bv_{m+1},\nu)=0, \quad \nu=L_2(\Omega). \label{9.24}
\end{gather}
Here $\rho_m$ is a positive constant.
\begin{th}
Suppose that conditions \eqref{4.6}, \eqref{4.7}, $(C4a)$ are satisfied, and 
let the function $\psi_1$ meets $(C0)$ and one of the conditions $(C1a)$,
$(C2a)$. Assume that $(v_0,p_0)$ is an arbitrary pair in $X\times L_2(\Omega)$.
Then, for an arbitrary $m$ there exists a unique pair $(v_{m+1},p_{m+1})$
satisfying \eqref{9.22}--\eqref{9.24}. Moreover, if
\begin{equation}\label{9.25}
0<\inf_{m\in\N}\rho_m\le\sup_{m\in\N}\rho_m<2r, 
\end{equation}
then
\begin{equation}\label{9.26}
v_m\to v_\lambda \mbox{ in } X, \qquad p_m\to p_\lambda \mbox{ in } L_2
(\Omega),
\end{equation}
where $(v_\lambda,p_\lambda)$ is the solution of \eqref{8.11}--\eqref{8.13}.
\end{th}
{\it Proof.} We set
\begin{equation}\label{9.27}
u_m=v_m-v_\lambda, \qquad q_m=p_m-p_\lambda.
\end{equation}
By subtracting \eqref{8.12} from \eqref{9.23} and \eqref{8.13} from \eqref{9.24},
we get
\begin{gather}
(L_3(v_{m+1})-L_3(v_\lambda),h)+r(Bu_{m+1},Bh)=(B^*q_m,h), \quad h\in X,
\label{9.28}\\
(q_{m+1}-q_m,\nu)=-\rho_m(Bu_{m+1},\nu), \quad \nu\in L_2(\Omega). \label{9.29}
\end{gather}
Taking $\nu=2q_{m+1}$ in \eqref{9.29}, we obtain
\begin{equation}\label{9.30}
\|q_{m+1}\|_{L_2(\Omega)}^2\,-\|q_m\|_{L_2(\Omega)}^2\,+\|q_{m+1}-q_m\|_{L_2
(\Omega)}^2=-2\rho_m(Bu_{m+1},q_{m+1}). 
\end{equation}
Take $h=u_{m+1}$ in \eqref{9.28}. Then \eqref{9.28}--\eqref{9.30} give
\begin{gather}
\|q_{m+1}\|_{L_2(\Omega)}^2-\|q_m\|_{L_2(\Omega)}^2+\|q_{m+1}-q_m\|_{L_2
(\Omega)}^2 \notag \\
+2\rho_m(L_3(v_{m+1})-(L_3(v_\lambda),u_{m+1})+2\rho_m\,r\|Bu_{m+1}
\|_{L_2(\Omega)}^2\notag \\
=-2\rho_m(Bu_{m+1},q_{m+1}-q_m). \label{9.31}
\end{gather}
\eqref{9.31} and Lemma 3.2 (see \eqref{3.15}) imply
\begin{gather}
\|q_{m+1}\|_{L_2(\Omega)}^2-\|q_m\|_{L_2(\Omega)}^2+\|q_{m+1}-q_m\|_{L_2
(\Omega)}^2 \notag\\
+4\rho_m a_3\int_\Omega [I(\tilde u+v_{m+1})^{\frac12}-I(\tilde u+v_\lambda)^
{\frac12}]^2\,dx \notag\\
+2\rho_mr\|Bu_{m+1}\|_{L_2(\Omega)}^2\le 2\rho_m\|Bu_{m+1}\|_{L_2(\Omega)}
\|q_{m+1}-q_m\|_{L_2(\Omega)}. \label{9.32}
\end{gather}
By applying the inequality $2ab\le a^2+b^2$ to the right-hand side of \eqref
{9.32}, we obtain
\begin{gather}
\|q_{m+1}\|_{L_2(\Omega)}^2-\|q_m\|_{L_2(\Omega)}^2+4\rho_m a_3\int_\Omega[I
(\tilde u+v_{m+1})^{\frac12}-I(\tilde u+v_\lambda)^{\frac12}]^2\,dx \notag\\
+\rho_m(2r-\rho_m)\|Bu_{m+1}\|^2_{L_2(\Omega)}\le 0. \label{9.33}
\end{gather}
By virtue of \eqref{9.25} there exists $\delta>0$ such that $\rho_m(2r-\rho_m)
\ge\delta$ for any $m$, and by \eqref{9.33}, $\|q_m\|_{L_2(\Omega)}\ge
\|q_{m+1}\|_{L_2(\Omega)}$. Thus, the sequence $\{\|q_m\|_{L_2(\Omega)}^2\}$
converges, i.e. $\lim \|q_m\|_{L_2(\Omega)}^2=\alpha\ge 0$, and \eqref{9.33}
yields
\begin{gather}
\int_\Omega[I(\tilde u+v_{m+1})^{\frac12}-I(\tilde u+v_\lambda)^{\frac12}]^2
\,dx \to 0, \label{9.34}\\
Bu_m\to 0 \mbox{ in } L_2(\Omega). \label{9.35}
\end{gather}
Since the function $u\to\int_\Omega u^2\,dx$ is a continuous mapping from $L_2
(\Omega)$ into $\R$ we obtain from \eqref{9.34} that
\begin{equation}\label{9.36}
\int_\Omega I(\tilde u+v_m)dx\to\int_\Omega I(\tilde u+v_\lambda)dx.
\end{equation}
Therefore
\begin{equation}\label{9.37}
\|v_m\|_X\le c, \qquad m\in \N,
\end{equation}
and by \eqref{9.23}, \eqref{3.37} we get $\|p_m\|_{L_2(\Omega)}\le c$ for all 
$m$. Therefore, a subsequence $\{v_\eta,p_\eta\}$ can be extracted such that
$v_\eta\rightharpoonup v_0$ in $X$, $p_\eta\rightharpoonup p_0$ in $L_2(\Omega)$.
 We pass to the limit by analogy with the above. Then we get $v_0=v_\lambda$,
$p_0=p_\lambda$. Due to \eqref{9.36} and by the uniqueness of the solution of
\eqref{8.11}--\eqref{8.13}, we obtain by analogy with the above (see \eqref
{9.13}) that
\begin{equation}\label{9.38}
v_m\to v_\lambda \mbox{ in } X.
\end{equation}
It follows from \eqref{9.28} and \eqref{3.37} that
\begin{equation}\notag
\|L_3(v_{m+1})-L_3(v_\lambda)+rB^*Bu_{m+1}\|_{X^*}=\|B^*q_m\|_{X^*}\ge\beta_1
\|q_m\|_{L_2(\Omega)}.
\end{equation}
This inequality, together with \eqref{9.35} and \eqref{9.38}, yields $q_m\to 0$
in $L_2(\Omega)$. Therefore \eqref{9.26} holds true.
$\blacksquare$

\subsection{Solving a nonlinear problem.}
We consider two methods for solving the nonlinear problem \eqref{9.23}, namely
the Birger-Kachanov method and the contraction method. Both methods transform
a nonlinear problem into a sequence of linear problems. 

We consider the problem: find a function $u$  satisfying
\begin{equation}\label{9.39}
u\in X, \quad (L_3(u),h)+r(Bu,Bh)=(f,h), \quad \forall h\in X,
\end{equation}
where $f\in X^*$.

For an arbitrary $v\in X$ we define the operator $M(v)\in\cal L(X,X^*)$ as 
follows:
\begin{gather}
(M(v)w,h)=2\int_\Omega [e(\vert E\vert,x)(\lambda+I(\tilde u+v))^{-\frac12}
\eps_{ij}(\tilde u+w)\,\eps_{ij}(h) \notag\\
+\psi_1(I(\tilde u+v),\vert E\vert,x)\,\eps_{ij}(\tilde u+w)\,\eps_{ij}(h)]dx
+r(Bw,Bh). \label{9.40}
\end{gather}
The Birger-Kachanov method consists in constructing a sequence $\{u_m\}$ such
that
\begin{equation}\label{9.41}
u_{m+1}\in X, \quad (M(u_m)u_{m+1},h)=(f,h), \quad \forall h\in X.
\end{equation}
The conditions for the convergence of the Birger-Kachanov method in the general 
situation were established in \cite{17}.

From the known results \cite{17}, \cite{7}, \cite{11}, the next theorem follows.
\begin{th}
Suppose the conditions \eqref{4.6}, \eqref{4.7}, $(C4a)$ are satisfied. Assume 
that $\psi_1$ is a nonincreasing function meeting the conditions $(C0)$ and
$(C1a)$. Let also $\lambda>0$ an $u_0$ be an arbitrary element of $X$. Then for
any $m$ there exists a unique solution $v_{m+1}$ of the problem \eqref{9.41}
and $u_m\to u$ in $X$, where $u$ is the solution of \eqref{9.39}.
\end{th}   
Consider now the contraction method. Let $A$  be a linear continuous 
selfadjoint and coercive mapping from $X$ into $X^*$, i.e.
\begin{gather}
(Au,h)=(u,Ah), \quad \vert(Au,h)\vert\le b_1\|u\|_X\|h\|_X,
\quad u,h\in X, \notag\\
(Au,u)\ge b_2\|u\|_X^2, \quad u\in X, \label{9.42}
\end{gather}
where $b_1$, $b_2$ are positive constants. By \eqref{9.42} the expression
\begin{equation}\label{9.43}
\|u\|_1=(Au,u)^{\frac12}
\end{equation}
defines that norm in $X$ that is equivalent to the norm $\|.\|_X$ and to the 
norm of $H^1(\Omega)^n$. By $X_1$ we denote the space $X$ equipped with the 
scalar product
\begin{equation}\label{9.44}
(u,h)_{X_1}=(Au,h)
\end{equation}
and with the norm \eqref{9.43}.

We study the following iterative method:
\begin{equation}\label{9.45}
u_{m+1}\in X, \quad (Au_{m+1},h)=(Au_m,h) 
-t[L_3(u_m,h)+r(Bu_m,Bh)-(f,h)], \quad \forall h\in X, 
\end{equation}
where $t$ is a positive constant.

We may define the operator $A$ by
\begin{equation}\label{9.46}
(Au,h)=\int_\Omega g\frac{\di u_i}{\di x_j}\,\frac{\di h_i}{\di x_j}\,dx,
\quad u,h\in X,
\end{equation}
where $g\in C(\overline \Omega)$, $g(x)\ge c_0>0$ for all $x\in\overline\Omega$.
Then, taking $h=(h_1,0)$ and $h=(0,h_2)$ in the case that $\Omega\subset\R^2$,
and $h=(h_1,0,0)$, $h=(0,h_2,0)$, $h=(0,0,h_3)$ for $\Omega\subset\R^3$, where
$h_i$ are arbitrary functions from $U=\{\eta\in H^1(\Omega)$, $\eta\big\vert_{S_1}=0
\}$, we split the problem \eqref{9.45} and obtain independent problems
for calculation $u_{mi}$, $i=1,\dots,n$. Such a split is very convenient for 
computations.

\begin{lem}
Suppose the conditions \eqref{4.6}, $(C0)$, $(C1a)$, $(C4a)$,  \eqref{9.42}
are satisfied. Then
\begin{gather}
(L_3(v)-L_3(h),v-h)\ge q_1\|v-h\|_1^2, \label{9.47}\\
\|L_3(v)-L_3(h)\|_{X_1^*}\le q_2\|v-h\|_1, \label{9.48}
\end{gather}
where
\begin{gather}
q_1=\mu_1b_1^{-1}, \notag\\
q_2=(\mu_2+4a_5\lambda^{-\frac12})b_2^{-\frac12}. \label{9.49}
\end{gather}
$\mu_1$, $\mu_2$ are defined by \eqref{3.8}.
\end{lem}
{\it Proof}. Taking into account the inequalities of \eqref{9.42}, and applying
Lemmas 3.1 and 3.4, we obtain  \eqref{9.47}, \eqref{9.48} with $q_1$, $q_2$
defined by \eqref{9.49}.
\begin{th}
Suppose the conditions \eqref{4.6}, $(C0)$, $(C1a)$, $(C4a)$ are satisfied. Let
$f\in X^*$ and the operator $A\in \cal L(X,X^*)$ meets \eqref{9.42}. Let also
$\lambda>0$ and $u_0$ be an arbitrary element of $X$. Then for $t\in (0,2\,q_1
\,q_3^{-2})$, where 
\begin{equation}\label{9.50}
q_3=q_2+r\|B^*B\|_{\cal L(X_1,X_1^*)},
\end{equation}
and for any $m$ there exists a unique solution $u_{m+1}$ of problem \eqref{9.45}
and the following estimate holds
\begin{equation}\label{9.51}
\|u_m-u\|_1\le\frac{k(t)^m}{1-k(t)}\,\|L_3(u_0)+rB^*B\,u_0-f\|_{X_1^*},
\end{equation}
where
\begin{equation}\label{9.52}
k(t)=(1-2q_1t+q_3^2\,t^2)^{\frac12}<1,
\end{equation}
and $u$ is the solution of \eqref{9.39}.

The function $k$ takes its minimal value $k(t_0)=(1-q_1^2\,q_3^{-2})^{\frac12}$
at the point $t_0=q_1\,q_3^{-2}$.
\end{th}
{\it Proof}. 
Let $N=L_3+rB^*B$. By Lemma 9.1 we have
\begin{gather}
(N(v)-N(h),h-v)\ge q_1\|v-h\|_1^2,  \notag\\
\|N(v)-N(h)\|_{X_1^*}\le q_3\|v-h\|_1, \quad v,h\in X, \label{9.53}
\end{gather}
where $q_3=q_2+r\|B^*B\|_{\cal L(X_1,X_1^*)}$.

Denote by $J$ the Riesz operator $J\in\cal L(X_1^*,X_1)$ that is defined as 
follows
\begin{equation}\label{9.54}
(Jg,h)_{X_1}=(g,h), \quad g\in X_1^*, \quad h\in X_1, \quad \|Jg\|_1=\|g\|_
{X_1^*}. 
\end{equation}
It is obvious that the problem \eqref{9.39} is equivalent to finding a fixed 
point $u=U_t(u)$, where $U_t:X_1\to X_1$,
\begin{equation}\label{9.55}
U_t(h)=h-tJ(N(h)-f).
\end{equation}
By \eqref{9.53}--\eqref{9.55} we have
\begin{gather}
\|U_t(v)-U_t(h)\|_1^2=\|v-h-tJ(N(v)-N(h))\|_1^2   \notag\\
=\|v-h\|_1^2-2t(N(v)-N(h),v-h)+t^2\|N(v)-N(h)\|_{X_1^*}^2  \notag\\
\le\|v-h\|_1^2-2tq_1\|v-h\|_1^2+t^2q_3^2\|v-h\|_1^2=(k(t)\|v-h\|_1)^2, \notag
\end{gather}
where $k(t)$ is defined by \eqref{9.52} and $k(t)<1$, if $t\in(0,2q_1q_3^{-2})$.
Therefore the mapping $U_t$ is a contraction, and the existence of a unique
solution $u$ of problem \eqref{9.39} and estimate \eqref{9.51} follow from the 
fixed point theorem (see e.q. \cite{14}).
$\blacksquare$.

{\bf Remark 9.1}. Equalities \eqref{9.49} and \eqref{9.50} imply that $q_3\to
\infty$ as $\lambda\to 0$. Therefore $k(t_0)=(1-q_1^2\,q_3^{-2})^{\frac12}$ 
the minimal value of $k(t)$ tends to unit as $\lambda\to 0$, and the iterative
method \eqref{9.45} provides slow convergence at small value of $\lambda$. The
reason of this is that the differentiable functional $Y_\lambda$ tends to 
nondifferentiable functional $Y$ (see \eqref{8.8}, \eqref{8.9}) as $\lambda
\to 0$.

\subsection{Solving the problem \eqref{5.8}--\eqref{5.10}.}
For the case that the conditions of Theorem 5.2 are satisfied, the operators
$\frac{\di J_\lambda}{\di h}$ and $L$ are strictly monotone and  Lipschitz continuous.
Therefore the algorithm of the augmented Lagrangian (see \eqref{9.22}--\eqref
{9.24}) can be used for the solution of the problem \eqref{5.8}--\eqref{5.10},
and the corresponding nonlinear systems can be solved by the Birger-Kachanov
method and by the contraction method.

Let us consider the general case that the conditions of Theorem 5.2 are not
satisfied. Let $\{\chi_i\}_{i=1}^{k_1(m)}$, $\{\eta_i\}_{i=1}^{k_2(m)}$ be bases 
in the spaces $X_m$ and $N_m$, respectively. Let also $k=k_1(m)+k_2(m)$. Define
a mapping $\cal M:\R^k\to\R^k$ as follows:
\begin{equation}\label{9.56}
\R^k\ni c=\{c_i\}_{i=1}^k \to\cal M(c)=\{\cal M_i(c)\}_{i=1}^k,
\end{equation}
where
\begin{gather}
\cal M_i(c)=\Big(\frac{\di J_\lambda}{\di h}\Big(\sum_{j=1}^{k_1(m)}c_j\chi_j,
\sum_{j=1}^{k_1(m)}c_j\chi_j\Big),\chi_i\Big)+\Big(L\Big(\sum_{j=1}
^{k_1(m)}c_j\chi_j\Big),\chi_i\Big) \notag\\
-\Big(B_m^*\sum_{j=k_1(m)+1}^{k}c_j\eta_{j-k_1(m)},\chi_i\Big)-(K+F,\chi_i),
\quad i=1,\dots,k_1(m). \label{9.57}\\
\cal M_i(c)=\Big(B_m\sum_{j=1}^{k_1(m)}c_j\chi_j,\eta_{i-k_1(m)}\Big),
\quad i=k_1(m)+1,\dots,k. \label{9.58}
\end{gather}
It is obvious that the problem \eqref{5.8}--\eqref{5.10} is equivalent to the
following one: find $\tilde c=(\tilde c_1,\dots,\tilde c_k)$ such that
\begin{equation}\label{9.59}
\tilde c\in\R^k, \qquad \cal M(\tilde c)=0.
\end{equation}
Define the functional
\begin{equation}\label{9.60}
\Phi_2(c)=\sum_{i=1}^k(\cal M_i(c))^2, \qquad c\in\R^k.
\end{equation}
The problem \eqref{9.59} is equivalent to the following one:
\begin{equation}\label{9.61}
\tilde c\in\R^k, \qquad \Phi_2(\tilde c)=\min_{c\in\R^k}\Phi_2(c)=0.
\end{equation}
In the case that the functions $\psi$ and $b$ are continuously differentiable 
the functional $\Phi_2$ is continuously differentiable in $\R^k$, and gradient 
method can be applied for calculation of a solution of the problem \eqref{9.61}.
Derivative of the mapping $v\to\frac{\di J_\lambda}{\di h}(v,v)+L(v)$ is 
defined in 5.2 (see \eqref{5.43}--\eqref{5.48}).

\section{Stationary problem with consideration for the inertia forces. 
Nonhomogeneous problem.}
\subsection{Basic equations and auxiliary results.}

The equations of motion with regard for the inertia forces read as follows:
\begin{equation}\label{10.1}
\rho u_j\,\frac{\di u_i}{\di x_j}+\,\frac{\di p}{\di x_i}\,-2\,\frac{\di}
{\di x_j}[\phi(I(u),\vert E\vert,\mu(u,E))\eps_{ij}(u)]=K_i \mbox{ in }\Omega,\quad
i=1,\dots,n.
\end{equation}
The condition of incompressibility is
\begin{equation}\label{10.2}
\diver u=0.
\end{equation}
We assume that velocities are specified on the boundary $S$ of $\Omega$, i.e.
\begin{equation}\label{10.3}
u\Big\vert_S=\hat u.
\end{equation}
We assume also
\begin{description}
\item[(C7)]
$\Omega$ is a bounded domain in $\R^n$, $n=2$ or $3$.
The boundary $S$ of $\Omega$ belongs to the class $C^2$ 
and consists of $l$ connected components $\Gamma_1,\dots,\Gamma_l$
$(l\ge 1)$,
\end{description}
and suppose that
\begin{equation}\label{10.4}
\hat u\in H^{\frac12}(S),\int_{\Gamma_i}\hat u_i\nu_ids=0, \quad i=1,\dots,l,
                \end{equation}
and the function $\phi$ is defined by \eqref{2.12a}.

The following lemma follows from the known results (see e.g. \cite{18,19}).

\begin{lem}
Suppose that the conditions $(C7)$ and \eqref{10.4} are satisfied. Then there 
exists a function $\tilde u$ such that
\begin{equation}\label{10.5}
\tilde u=\rot\eta,\quad \eta=(\eta_1,\dots,\eta_n)\in H^2(\Omega)^n,
\quad \tilde u\big\vert_S=\hat u,
\end{equation}
moreover for an arbitrary $\alpha>0$ one can choose a vector-valued function
$\eta$ such that
\begin{equation}\label{10.6}
\|\tilde u_i v_j\|_{L_2(\Omega)}\le \alpha\|v\|_X,\quad v\in H^1_0(\Omega)^n,
\quad i,j=1,\dots,n.
\end{equation}
\end{lem}
We set
\begin{equation}\label{10.7}
q(u,v,w)=\rho\int_\Omega u_k\frac{\di v_i}{\di x_k}w_idx, \qquad 
u,v,w\in H^1(\Omega)^n.
\end{equation}
Obviously
\begin{equation}\label{10.8}
\vert q(u,v,w)\vert\le\rho\sum_{i,k=1}^n\|u_k\|_{L_4(\Omega)}\Big\|\frac{\di
v_i}{\di x_k}\Big\|_{L_2(\Omega)}\|w_i\|_{L_4(\Omega)},
\end{equation}
Therefore, the trilinear form $q$ is continuous in $H^1(\Omega)^n\times H^1
(\Omega)^n \times H^1(\Omega)^n$.

We consider the following spaces
\begin{gather}
\cal X=H_0^1(\Omega)^n\quad\mbox{with the norm } \|\cdot\|_{\cal X}=\|\cdot\|
_X, \label{10.9} \\
\cal V=\{w\in \cal X,\,\,\diver w=0\}\quad \mbox{with the norm } \|\cdot\|_
{\cal V}=\|\cdot\|_X, \label{10.10}\\
\cal N=\{w\in L_2(\Omega), \,\,\int_\Omega w\,dx=0\}\quad \mbox{with the norm }
\|\cdot\|_{\cal N}=\|\cdot\|{L_2(\Omega)}. \label{10.11}
\end{gather}
It is easy to verify that
\begin{equation}\label{10.12}
q(z,w,h)=-q(z,h,w),\quad z\in\cal V, \quad w,h\in H^1(\Omega)^n, \quad 
n=2 \mbox{ or }3,\qquad q(z,h,h)=0.
\end{equation}
Define a trilinear form $q_1$ as follows:
\begin{equation}\label{10.13}
q_1(v,w,h)=\frac12\,q(v,w,h)-\frac12\,q(v,h,w), \quad v,h,w\in H^1(\Omega)^n.
\end{equation}
It is evident that
\begin {alignat}{3}
&q_1(v,h,h)=0,           &\qquad &v,h\in H^1(\Omega)^n, \label{10.14}\\
&q_1(v,w,h)=-q_1(v,h,w), &\qquad &v,h,w\in H^1(\Omega)^n. \label{10.15}
\end {alignat}
\begin{lem}
Suppose that
\begin{align}
&v_k\rightharpoonup v\mbox{ in }\cal X, \label{10.16}\\
&v_k\to v \mbox{ in }L_4(\Omega)^n. \label{10.17}
\end{align}
Then, for an arbitrary fixed $h\in H^1_0(\Omega)^n$ the following relations 
hold
\begin {align}
&q_1(v_k,v_k,h)      \to q_1(v,v,h), \label{10.18}\\
&q_1(\tilde u,v_k,h) \to q_1(\tilde u,v,h), \label{10.19}\\
&q_1(v_k,\tilde u,h) \to q_1(v,\tilde u,h). \label{10.20}
\end {align}
In this case if $v\in\cal V$, then
\begin{equation} \label{10.21}
q_1(v,v,h)=q(v,v,h), \quad q_1(\tilde u,v,h)=q(\tilde u,v,h), \quad
q_1(v,\tilde u,h)=q(v,\tilde u,h).
\end{equation}
\end{lem}
{\bf Proof.} We have
\begin{equation}\label{10.22}
\int_\Omega(v_{kj}h_i-v_j h_i)^2\,dx\le\Big(\int_\Omega(v_{kj}-v_j)^4\,dx
\Big)^{\frac12}\Big(\int_\Omega h_i^4\,dx\Big)^{\frac12},
\end{equation}
where $v_{kj}$ are the components of the vector-valued function $v_k$. By
\eqref{10.17} the left-hand side of \eqref{10.22} tends to zero.

Therefore,
\begin{equation}\label{10.23}
v_{kj}h_i\to v_j h_i \mbox{ in } L_2(\Omega) \mbox{ as } k\to\infty, \quad
i,j=1,\dots,n.
\end{equation}
\eqref{10.16} and \eqref{10.23} yield
\begin{equation}\label{10.24}
q(v_k,v_k,h)\to q(v,v,h).
\end{equation}
Application Green's formula gives
\begin{equation}\notag
q(v_k,h,v_k)=\rho\int_\Omega v_{kj}\frac{\di h_i}{\di x_j}\,v_{ki}\,dx=-\rho
\int_\Omega h_i\frac{\di}{\di x_j}(v_{kj}v_{ki})dx,
\end{equation}
and by analogy with the stated above we obtain
\begin{equation}\notag
q(v_k,h,v_k)\to q(v,h,v).
\end{equation}
Therefore \eqref{10.18} holds. It is evident that \eqref{10.19}, \eqref{10.20}
follows from \eqref{10.16} and \eqref{10.17}.

In the special case that $v\in\cal V$ we have
\begin{equation}\label{10.25}
q(v,v,h)=\rho\int_\Omega v_j\frac{\di v_i}{\di x_j}\,h_i\,dx=-\rho\int_\Omega 
v_j\frac{\di h_i}{\di x_j}\,v_i\,dx=-q(v,h,v). 
\end{equation}
\eqref{10.13} and \eqref{10.25} imply \eqref{10.21}.
\begin{lem}
Suppose the conditions $(C7)$ and \eqref{10.4} are satisfied. Then for an 
arbitrary $\xi>0$ one can choose a vector valued function $\eta$ such that
\eqref{10.5} is satisfied and in addition
\begin{equation}\label{10.26}
\vert q_1(v,\tilde u,v)\vert\le \xi\|v\|_X^2, \qquad v\in\cal X. 
\end{equation}
\end{lem}
{\bf Proof}. By Green's formula we obtain
\begin{gather}
q_1(v,\tilde u,v)=\frac12\rho\,\int_\Omega\Big[v_j\frac{\di\tilde u_i}{\di x_j}
\,v_i-v_j\frac{\di v_i}{\di x_j}\tilde u_i\Big]dx \notag\\
=\frac12\,\rho\int_\Omega\Big[\Big(-\sum_{j=1}^n\frac{\di v_j}{\di x_j}\Big)
(\tilde u_i\,v_i)-2\frac{\di v_i}{\di x_j}\tilde u_i v_j\Big]dx. \qquad 
v\in\cal X.\notag
\end{gather}
It follows from here that
\begin{equation}\notag
\vert q_1(v,\tilde u,v)\vert\le c_1\|v\|_X\,\sum_{i,j=1}^n\|\tilde u_i v_j\|_
{L_2(\Omega)},
\end{equation}
and Lemma 10.3 follows from this inequality and Lemmas 10.1. 

\subsection{Boundary value problem.}
We consider the problem: find $v$ satisfying
\begin{align}
&v\in\cal V \label{10.27}\\
&\Big(\frac{\di J_\lambda}{\di h}(v,v),w\Big)+(L(v),w)+q(v,v,w)+q(\tilde u,v,
w)+q(v,\tilde u,w)-(y,w)=0, \quad w\in\cal V, \label{10.28}
\end{align}
where
\begin{equation}\label{10.29}
(y,w)=\int_\Omega K_i w_i \,dx-\int_\Omega\tilde u_j\frac{\di\tilde u_i}{\di x_j}
\,w_i\,dx. 
\end{equation}
It follows from \eqref{4.7} and \eqref{10.5} that $y\in\cal X^*$. The left-hand
 side of \eqref{10.28} belongs to the polar set
\begin{equation}\notag
\cal V^{\circ}=\{f\in\cal X^*, \quad (f,w)=0, \quad w\in\cal V\}. 
\end{equation}
Therefore, there exists a function $p\in\cal N$ such that the pair $(v,p)$ is a
solution of the following problem:
\begin{gather}
(v,p)\in \cal X\times\cal N, \label{10.30}\\
\Big(\frac{\di J_\lambda}{\di h}(v,v),w\Big)+(L(v),w)+q(v,v,w)+q(\tilde u,v,w)
\notag\\
+q(v,\tilde u,w)-(B^*p,w)=(y,w), \quad w\in\cal X, \label{10.31}\\
(Bv,\gamma)=0, \quad \gamma\in\cal N. \label{10.32}
\end{gather}
By use of Green's formula it can be seen that, if $(v,p)$ is a solution of 
problem \eqref{10.30}--\eqref{10.32}, then $(u,p)$ with $u=\tilde u+v$ is a 
solution of problem \eqref{10.1}--\eqref{10.3} in the sense of distributions. 
On the contrary, if $(u,p)$ is a classical solution of problem 
\eqref{10.1}--\eqref{10.3}, then the pair $(v,p)$ with $v=u-\tilde u$ is a 
solution of problem \eqref{10.30}--\eqref{10.32}.

Let $\{\cal X_m\}$, $\{\cal N_m\}$ be sequences of finite-dimensional subspaces
in $\cal X$ and $\cal N$ which satisfy the following conditions
\begin{alignat}{3}
&\lim_{m\to\infty}\,\inf_{z\in\cal X_m}\,\|w-z\|_X=0, &\quad &w\in\cal X,
\label{10.33}\\
&\lim_{m\to\infty}\,\inf_{y\in\cal N_m}\,\|h-y\|_{L_2(\Omega)}=0, &\quad
&h\in\cal N, \label{10.34}\\
&\inf_{\mu\in\cal N_m}\,\sup_{w\in\cal X_m}\,\frac{(B_mw,\mu)}{\|w\|_X\|\mu\|_
{L_2(\Omega)}}\ge\beta>0, &\quad &m\in\N, \label{10.35}\\
&\qquad \cal X_m\subset\cal X_{m+1},\qquad \cal N_m\subset\cal N_{m+1}.&\qquad
&\qquad \label{10.36}
\end{alignat}
We introduce the spaces $\cal V_m$ and $\cal V_m^\circ$ by
\begin{alignat}{5}
&\cal V_m=\{u\in\cal X_m, &\quad &(B_mu,\gamma)=0, &\quad &\gamma\in\cal N_m\}, 
\label{10.37}\\
&\cal V_m^\circ=\{q\in\cal X_m^*, &\quad &(q,u)=0, &\quad &u\in\cal V_m\}. 
\label{10.38}
\end{alignat}
Define approximate solutions of problem \eqref{10.30}--\eqref{10.32} as follows:
\begin{gather}
(v_m,p_m)\in\cal X_m\times\cal N_m, \label{10.39}\\ 
\Big(\frac{\di J_\lambda}{\di h}(v_m,v_m),w\Big)+(L(v_m),w)+q_1(v_m,v_m,w)+q_1
(\tilde u,v_m,w) \notag\\
+q_1(v_m,\tilde u,w)-(B_m^* p_m,w)=(y,w), \qquad w\in\cal X_m, \label{10.40}\\
(B_m,v_m,\gamma)=0, \qquad \gamma\in\cal N_m. \label{10.41} 
\end{gather}
\begin{th}
Suppose that $K\in L_2(\Omega)^n$ and the condition $(C4)$ is satisfied. Let the 
function $\psi$ meets one of conditions $(C1)$, $(C2)$, $(C3)$ ($\phi$ 
replaced by $\psi$). Assume that $(C7)$, \eqref{10.4}, \eqref{10.5} and 
\eqref{10.26} with  $\xi\le a_1$ are fulfilled. Let also $\{\cal X_m\}$
and $\{\cal N_m\}$ be sequences of finite-dimensional subspaces in $\cal X$ and
$\cal N$ respectively, such that \eqref{10.33}--\eqref{10.36} hold.
Then, for an arbitrary fixed $\lambda>0$ and an arbitrary $m\in\N$ there exists
a solution $(v_m,p_m)$ of problem \eqref{10.39}--\eqref{10.41}, and a 
subsequence $\{v_k,p_k\}$ can be extracted from the sequence $\{v_m,p_m\}$ such
that $v_k\rightharpoonup v$ in $\cal X$, $p_k\rightharpoonup p$ in $\cal N$,
where $v,p$ is a solution of the problem \eqref{10.30}--\eqref{10.32}.
\end{th}
{\bf Proof}. Define an operator $M:\cal X\to\cal X^*$ by
\begin{equation}\label{10.42}
(M(g),w)=\Big(\frac{\di J_\lambda}{\di h}(g,g),w\Big)+(L(g),w)+q_1(g,q,w)
+q_1(\tilde u,g,w)+q_1(g,\tilde u,w), \quad g,w\in\cal X.
\end{equation}
It follows from \eqref{10.39}--\eqref{10.42} that $v_m$ is a solution of the 
problem
\begin{equation}\label{10.43}
v_m\in\cal V_m,\quad (M(v_m),w)=(y,w), \quad w\in\cal V_m. 
\end{equation}
Bearing in mind \eqref{10.14} and that $\xi\le a_1$ in \eqref{10.26} we
obtain by analogy with the proof of Theorem 5.1 (see \eqref{5.13}, \eqref{5.16},
that
\begin{equation}\label{10.44}
z(e)=(M(e),e)-(y,e)\ge a_1 \|e\|_X^2-c\|e\|_X, \quad e\in\cal X,\,\,
\lambda>0.
\end{equation}
Therefore, $z(e)\ge0$ for $\|e\|_X\ge r\,=\frac{c}{a_1}$, and there exists a 
solution of \eqref{10.43} with
\begin{equation}\label{10.45}
\|v_m\|_X\le r, \quad \|M(v_m)\|_{\cal X^*}\,\le c_1 \quad m\in\N.
\end{equation}
For an arbitrary $f\in\cal X^*$ we denote by $Gf$ the restriction of $f$ to 
$\cal X_m$. Then $Gf\in\cal X_m^*$, and by \eqref{10.43} we obtain
\begin{equation}\label{10.46}
G(M(v_m)-y)\in\cal V_m^\circ.
\end{equation}
Therefore, there exists a unique $p_m\in\cal N_m$ such that \eqref{10.40} is
satisfied (see Lemma 3.6), and \eqref{10.35} yields
\begin{equation}\label{10.47}
\|p_m\|_{\cal N}\le c_2.
\end{equation}
By \eqref{10.45}, \eqref{10.47} we can extract a subsequence $\{v_k,p_k\}$ such
that
\begin{align}
v_k    &\rightharpoonup v \mbox{ in } \cal X, \label{10.48}\\
v_k    &\to v \mbox{ in } L_4(\Omega)^n \mbox{ and a.e. in } \Omega, \label{10.49}\\
M(v_k) &\rightharpoonup\eta \mbox{ in } \cal X^*, \label{10.50}\\
p_k    &\rightharpoonup p \mbox{ in } \cal N. \label{10.51}
\end{align}
Next we use Lemma 10.2 and by analogy with the proof of Theorem 5.1 we pass to 
the limit in \eqref{10.40}, \eqref{10.41} with $m$ replaced by $k$, and obtain
\begin{gather}
(M(v),w)-(B^*\,p,w)=(y,w), \quad w\in\cal X. \label{10.52}\\
(Bv,\gamma)=0, \quad \gamma\in\cal N. \label{10.53}
\end{gather}
Taking into consideration that \eqref{10.21} is satisfied, for $v\in\cal V$,
we get that pair $(v,p)$ is a solution of the problem \eqref{10.30}--\eqref
{10.32}.

\section{Stationary problem with consideration for the inertia forces. Mixed
problem.}

\subsection{Formulation of the problem and an existence result.}
As before we consider that $S_1$ and $S_2$ are open subsets of the boundary 
$S$ of $\Omega$ such that $S_1$ is non-empty, $S_1\cap S_2=\emptyset$  and 
$\overline S_1\cup \overline S_2=S$. We study the problem on searching for a 
pair of functions $(u,p)$ which
satisfy the motion equations \eqref{10.1}, the condition of incompressibility
\eqref{10.2} and the mixed boundary conditions, wherein velocities are specified 
on $S_1$ and surface forces are given on $S_2$, i.e.
\begin{gather}
u\Big\vert_{S_1}=\hat u, \label{11.1}\\
[-p\delta_{ij}+2\phi(I(u),\vert E\vert,\mu(u,E))\eps_{ij}(u)]\nu_j\Big
\vert_{S_2}=F_i, \quad i=1,\dots,n. \label{11.2}
\end{gather}
It is obvious that in the special case that $S_2$ is an empty set this problem
transforms into the problem considered in Section 10.

We assume that $\phi$ is defined by \eqref{2.12a} and $\hat u\in H^{\frac12}(S_1)$.
Then there exists a function $\tilde u$ satisfying \eqref{4.6}.

Let us define operators $M_1:X\to X^*$, $M_2:X\to X^*$ and an element 
$\chi\in X^*$ as follows:
\begin{gather}
(M_1(w),g)=\Big(\frac{\di J_\lambda}{\di h}(w,w),g\Big)+(L(w),g), \quad w,g\in
X, \label{11.3}\\
(M_2(w),g)=q(\tilde u,w,g)+q(w,\tilde u,g), \quad w,g\in X; \label{11.4}\\
(\chi,g)=\int_\Omega K_i g_i\,dx+\int_{S_2}F_i g_i\,ds-q(\tilde u,\tilde u,g),
\quad g\in X. \label{11.5}
\end{gather}
We consider the problem
\begin{gather}
(v,p)\in X\times L_2(\Omega), \label{11.6}\\
(M_1(v),w)+(M_2(v),w)+q(v,v,w)-(B^* p,w)=(\chi,w), \quad w\in X, \label{11.7}\\
(Bv, \gamma)=0, \qquad \gamma\in L_2(\Omega). \label{11.8}
\end{gather}
By using Green's formula, one may show that, if $(v,p)$ is a solution of the
problem \eqref{11.6}--\eqref{11.8}, then $(u,p)$ with $u=\tilde u+v$ is a 
solution of the problem \eqref{10.1}, \eqref{10.2}, \eqref{11.1}, \eqref{11.2}
in the distribution sense. On the contrary, if $(u,p)$ is a solution of \eqref
{10.1}, \eqref{10.2}, \eqref{11.1}, \eqref{11.2} such that \eqref{11.6} holds 
with $v=u-\tilde u$, then $(v,p)$ is a solution of the problem \eqref{11.6}
--\eqref{11.8}.
Define also the following constants
\begin{align}
&r_1=\sup_{w\in V,\|w\|_X\le 1}\,q(w,w,w), \notag\\
&r_2=\inf_{w\in V,\|w\|_X\le 1}\,(M_2(w),w), \notag\\
&r_3=\sup_{w\in V,\|w\|_X\le 1}\,\vert (\chi,w)\vert. \label{11.9}
\end{align}
Consider the space
\begin{equation}\label{11.10}
P=\{w\in H^1(\Omega)^n, \qquad \diver w=0\}.
\end{equation}
The space $P$ is presented in the form $P=V\oplus V^{\perp}$, where $V$ is given
by \eqref{3.2} and $V^{\perp}$
is the orthogonal complement of $V$ in $P$. Evidently that the constants $r_2$
and $r_3$ in \eqref{11.9} depend on the function $\tilde u$ satisfying \eqref
{4.6}. 
Let $\tilde u_1$ be a function from $V^\perp$ that satisfies \eqref{4.6}. Then
the function $\tilde u=\tilde u_1+\tilde u_2$, where $\tilde u_2$ is an 
arbitrary element of $V$, meets \eqref{4.6}. We assume that there exists a 
function $\tilde u_2\in V$ such that the following inequalities hold
\begin{equation}\label{11.11}
r_4=a_1+\frac{r_2}2>0, \qquad r_4^2>r_1 r_3,
\end{equation}
where $a_1$ is the positive constants from \eqref{2.16}. It is evident that
\eqref{11.11} holds, if the norms of the functions $K,F$, and $\tilde u$  are 
not large.
\begin{lem}
Suppose the condition $(C4)$ is satisfied and the function $\psi$ meets one of the 
conditions $(C1)$, $(C2)$, $(C3)$ ($\phi$ replaced by $\psi$). Let also \eqref
{11.11}  holds.

Then the following inequality is valid:
\begin{gather}
\beta(w)=(M_1(w),w)+(M_2(w),w)+q(w,w,w)-(\chi,w)\ge 0, \notag\\
\mbox{if } w\in V \quad \mbox{and}\quad \|w\|_X=\frac{r_4-\sqrt{r_4^2-r_1 
r_3}}{r_1}\,=\xi. \label{11.12}
\end{gather}
\end{lem}
{\bf Proof}. It follows from \eqref{11.9} and \eqref{11.12} that
\begin{equation}\label{11.13}
\beta(w)\ge\beta_1(\|w\|_X)=(2a_1+r_2)\|w\|_X^2-r_1\|w\|_X^3-r_3\|w\|_X,\quad
w\in V. 
\end{equation}
Consider the quadratic equation
\begin{equation}\notag
2r_4 y-r_1 y^2-r_3=0,
\end{equation}
$r_4$ being defined in \eqref{11.11}. Its roots are those of the equation 
$\beta_1(y)=0$ and they are equal to
\begin{equation}\notag
y_1=\frac{r_4-\sqrt{r_4^2-r_1 r_3}}{r_1}, \qquad y_2=\frac{r_4+\sqrt{r_4^2-r_1 
r_3}}{r_1}.
\end{equation}
If \eqref{11.11} holds, then $y_1$ and $y_2$ are real and $\beta_1(y)\ge 0$ for 
$y\in[y_1, y_2]$. Therefore \eqref{11.11} yields \eqref{11.12}, and the lemma 
is proved.
$\blacksquare$
\begin{th}
Let $\Omega$ be a bounded domain in $\R^n$, $n=2$ or $3$ with a Lipschitz 
continuous boundary $S$. Suppose the condition $(C4)$ is satisfied and the 
function $\psi$ meets one of the conditions $(C1)$, $(C2)$, $(C3)$ ($\phi$
replaced by $\psi$). Let also \eqref{11.11} holds. Then, for an arbitrary 
$\lambda>0$ there exists a solution of the problem \eqref{11.6}--\eqref{11.8}.
\end{th}
{\bf Proof}. It follows from \eqref{11.6}--\eqref{11.8} that the function $v$
is a solution of the problem
\begin{gather}
v\in V, \notag\\
(M_1(v),w)+(M_2(v),w)+q(v,v,w)=(\chi,w), \quad w\in V. \label{11.14}
\end{gather}
Let $\{V_m\}$ be a sequence of finite-dimensional subspaces of $V$ such that
\begin{gather}
\lim_{m\to\infty}\,\inf_{z\in V_m}\,\|w-z\|_X=0, \qquad w\in V, \label{11.15}\\
V_m\subset V_{m+1}. \label{11.16}
\end{gather}
We search for the Galerkin approximations $v_k$ satisfying
\begin{gather}
v_m\in V_m, \notag\\
(M_1(v_m),w)+(M_2(v_m),w)+q(v_m,v_m,w)=(\chi,w), \qquad w\in V_m. 
\label{11.17}
\end{gather}
By virtue of Lemma 11.1 there exists a solution of the problem \eqref{11.17} and
$\|v_m\|_X\le\xi$. Thus we can extract a subsequence $\{v_k\}$ such that 
$v_k\rightharpoonup v_0$ in $V$. We pass to the limit as $k\to\infty$ in \eqref
{11.17}, with $m$ changed by $k$. In this case by analogy with the stated above,
(see the proofs of Theorems 5.1, 10.1 and Lemma 10.2), we use the methods of 
monotonicity and compactness. Thus, we get that the function $v=v_0$ is a 
solution of the problem \eqref{11.14}. Now from Lemma 3.5 it follows that there 
exists a function $p\in L_2(\Omega)$ such that the pair $(v,p)$ is a solution 
of the problem \eqref{11.6}--\eqref{11.8}.

\subsection{Approximation of the problem \eqref{11.6}--\eqref{11.8}.}
Let $\{X_m\}$, $\{N_m\}$ be sequences of finite dimensional subspaces of $X$ 
and $L_2(\Omega)$ which satisfy the conditions \eqref{3.38}, \eqref{3.39},
\eqref{3.43} and \eqref{5.11}. We search for an approximate solutions of the
problem \eqref{11.6}--\eqref{11.8} in the form
\begin{gather}
(v_m,p_m)\in X_m\times N_m, \label{11.18}\\
(M_1(v_m),w)+(M_2(v_m),w)+q(v_m,v_m,w)-(B_m^* p_m,w)=(\chi,w), 
\quad w\in X_m, \label{11.19}\\
(B_m v_m,\gamma)=0, \qquad \gamma\in N_m. \label{11.20}
\end{gather}
From the point of view of applications, in particular, of computation, the
problem \eqref{11.18}--\eqref{11.20} is considerably more preferable than
\eqref{11.17}. So, we study the question of convergence of the approximations
$\{v_m,p_m\}$.

Define constants $\eta_1-\eta_4$ by
\begin{align}
&\eta_1=\sup_m\,\max_{w\in d_m}\,\vert q(w,w,w)\vert, \notag\\
&\eta_2=\inf_m\,\min_{w\in d_m}\,(M_2(w),w), \notag\\
&\eta_3=\sup_m\,\max_{w\in d_m}\,\vert(\chi,w)\vert, \notag\\
&\eta_4=a_1+\frac{\eta_2}2, \label{11.21}
\end{align}
where 
\begin{equation}\label{11.22}
d_m=\{w\in X_m, \quad (B_m w,\gamma)=0, \quad \gamma\in N_m, \quad 
\|w\|_X\le 1\}.
\end{equation}
Note that $\eta_1\ge r_1$, $\eta_2\le r_2$, $\eta_3\ge r_3$ (see \eqref{3.38},
\eqref{3.39} and \eqref{11.9}).

Define a mapping $M_3$: $X\to X^*$ as follows:
\begin{equation}\label{11.23}
(M_3(v),w)=(M_2(v),w)+q(v,v,w)=q(\tilde u,v,w)+q(v,\tilde u,w)+q(v,v,w),
\quad v,w\in X.
\end{equation}
\begin{lem}
Suppose that \eqref{4.6} is satisfied and let
\begin{equation}\label{11.24}
\{v_k\}\subset X, \quad v_k\rightharpoonup v \text{ in } X.
\end{equation}
Then
\begin{align}
&\lim(M_3(v_k),w)=(M_3(v),w), \quad w\in X, \label{11.25}\\
&\lim(M_3(v_k),v_k)=(M_3(v),v). \label{11.26}
\end{align}
\end{lem}
{\bf Proof.} It follows from \eqref{11.24} that
\begin{equation}\label{11.27}
v_k\to v \text{ in } L_4(\Omega),
\end{equation}
and \eqref{11.25} arises from the proof of Lemma 10.2, see \eqref{10.24}.

We have
\begin{gather}
\vert q(v_k,v_k,v_k)-q(v,v,v)\vert=\Big\vert \int_\Omega(v_{kj}-v_j)\frac{\di v_
{ki}}{\di x_j}v_{ki}\,dx \notag\\
+\int_\Omega v_j\Big(\frac{\di v_{ki}}{\di x_j}-\frac{\di v_i}{\di x_j}\Big)\,
v_{ki}\,dx+\int_\Omega v_j\frac{\di v_i}{\di x_j}(v_{ki}-v_i)\,dx\Big\vert.
\label{11.28}
\end{gather}
\eqref{11.27} implies (see \eqref{10.22}) that
\begin{equation}\label{11.29}
v_j v_{ki}\to v_j v_i \text{ in } L_2(\Omega).
\end{equation}
By \eqref{11.24}, \eqref{11.27}, \eqref{11.29} each addend in the right-hand
side of the equality \eqref{11.28} tends to zero. Therefore,
\begin{equation}\notag
\lim q(v_k,v_k,v_k)=q(v,v,v)
\end{equation}
and \eqref{11.26} holds true.
$\blacksquare$

\begin{th}
Let $\Omega$ be a bounded domain in $\R^n$, $n=2$ or $3$ with a Lipschitz
continuous boundary $S$. Suppose the condition $(C4)$ is satisfied and the 
function $\psi$ meets one of the conditions $(C1)$, $(C2)$, $(C3)$ ($\phi$ 
replaced by $\psi$). Let also $\{X_m\}, $$\{N_m\}$ be sequences of finite 
dimensional subspaces of $X$ and $L_2(\Omega)$ which satisfy the conditions
\eqref{3.38}, \eqref{3.39}, \eqref{3.43} and \eqref{5.11}. Finally, assume that
\begin{equation}\label{11.30}
\eta_4>0, \qquad \eta_4^2>\eta_1\eta_3.
\end{equation}
Then, for an arbitrary fixed $\lambda>0$, and for each $m\in\N$, there exists a 
solution of the problem \eqref{11.18}--\eqref{11.20}, and a subsequence $\{(v_k,
p_k)\}$ can be extracted from the sequence $\{(v_m, p_m)\})$ such that $v_k
\rightharpoonup v$ in $X$, $p_k\rightharpoonup p$ in $L_2(\Omega)$, where $(v,p)$
is a solution of the problem \eqref{11.6}--\eqref{11.8}.
\end{th}
The proof of this theorem is analogous to that of Theorem 10.1, and it is not 
given because of this.

{\bf Remark 11.1}. In the general case we can consider that the density of an 
electrorheological fluid depends on the module of the vector of electric field
strength, i.e. $\rho=\rho(\vert E\vert)$, and
\begin{equation}\label{11.31}
\rho_2\ge\rho(y)\ge\rho_1, \quad y\in\R_+,
\end{equation}
where $\rho_1$, $\rho_2$ are positive constants. It is easy to see that all
 results of Sections 10 and 11 still stand valid in the case that $\rho$ is a 
function satisfying the condition \eqref{11.31}.

\end{document}